\documentclass[prb,showpacs,twocolumn]{revtex4}
\usepackage{amsmath,amssymb,epic,eepic}
\usepackage[dvips]{graphicx}
\usepackage{epsfig}
\usepackage{color}

\begin{document}

\title{Quantum detection of electronic flying qubits
in the integer quantum Hall regime}

\author{G.~F\`eve$^{1,2}$}
\author{P.~Degiovanni$^{3,4}$}
\author{Th.~Jolicoeur$^5$}

\affiliation{(1) Laboratoire Pierre Aigrain, D{\'e}partement de Physique
de l'Ecole Normale Sup\'erieure, 24 rue Lhomond, 75231 Paris Cedex
05, France}

\affiliation{(2) Laboratoire de Photonique et Nanostructures, CNRS,
Route de Nozay, 91460 Marcoussis, France}

\affiliation{(3) Universit\'e de Lyon, F\'ed\'eration de Physique Andr\'e Marie Amp\`ere,
CNRS - Laboratoire de Physique de l'Ecole Normale Sup\'erieure de Lyon,
46 All\'ee d'Italie, 69364 Lyon Cedex 07,
France}

\affiliation{(4) Department of Physics, Boston University, 590
Commonwealth avenue, Boston MA 02251, USA}

\affiliation{(5) Laboratoire de Physique Th\'eorique et Mod\`eles
Statistiques, Universit\'e Paris-Sud, 91405 Orsay Cedex, France}

%%%%%%%%%%%%%%%%%%%%%%%%%%%%%%%%%%%%%%%%%%%%%%%%%%%%%%%%%%%%%%%%%%%%%%%%%%%%%%%%%

\begin{abstract}
We consider a model of a detector of ballistic electrons at the edge
of a two-dimensional electron gas in the integer quantum Hall
regime. The electron is detected by capacitive coupling to a gate
which is also coupled to a passive $RC$ circuit. Using a quantum
description of this circuit, we determine the signal over noise
ratio of the detector in term of the detector characteristics. The
back-action of the detector on the incident wavepacket is then
computed using a Feynman-Vernon influence functional approach. Using
information theory, we define the appropriate notion of quantum
limit for such an "on the fly" detector. We show that our particular
detector can approach the quantum limit up to logarithms in the
ratio of the measurement time over the $RC$ relaxation time. We
argue that such a weak logarithmic effect is of no practical
significance. Finally we show that a two-electron interference
experiment can be used to probe the detector induced decoherence.
\end{abstract}


\maketitle

%%%%%%%%%%%%%%%%%%%%%%%%%%%%%%%%%%%%%%%%%%%%%%%%%%%%%%%%%%%%%%%%%%%%%%%%%%%%%%%%

\section{Introduction}

Among qubit implementations, solid state devices such as superconducting
qubits \cite{Makhlin:2001-1} or quantum dot qubits based on charge
\cite{Hu:2000-1} or spin degrees of freedom \cite{Loss:1998-1}
are especially promising because of their potential scalability. Were decoherence problems
solved, complex
circuits involving several qubits, quantum gates and detectors could in principle be manufactured
using standard nanofabrication techniques.
However, contrarily to recently proposed trapped ions\cite{Klelpinski:2002-1} and
Rydberg atom\cite{Mozley:2006-1}
architectures, solid state qubits cannot be displaced at will. Thus, any quantum computing architecture based
on solid state qubits requires a quantum bus efficiently transmitting quantum information
between various parts of the circuit \cite{DV}. An example of a quantum bus architecture is provided by
circuit QED in which superconducting qubits are coupled through one stationary mode
of a superconducting resonator \cite{Blais:2004-1}. Experimental realization of
circuit QED devices has been achieved recently~\cite{Wallraff:2004-1}.

Another example involves flying qubits based on electrons
propagating at the edge of a two dimensional electron gas (2DEG) in
the integer quantum Hall regime. These flying qubits have been proposed as the building block of a quantum
computation architecture \cite{Ionicioiu:2001-1} or to perform quantum state transfer
between charge or spin qubits \cite{Stace}.
In fact, in the presence of a quantizing magnetic field, edge modes
are an almost ideal chiral ballistic system without any backscattering
and with millimetric elastic mean free path.
Recently, the on-demand injection of an energy resolved electronic edge excitation has
been experimentally demonstrated \cite{Feve07}. Detection of single-electron
edge excitations is the next crucial step needed to build quantum Hall flying qubit
devices. Although detection of charge
density waves at the edge of a quantum Hall droplet has already been demonstrated
in an experiment by Ashoori et al.~\cite{Ashoori92},
it is desirable to extend such measurements to the single-electron regime for quantum computation
purposes.

In this paper, we present a model for the detection of
single-electron edge excitations of a 2DEG in the IQHE regime. Our
detector is based on a capacitive coupling between the 2DEG and a
mesoscopic gate connected to a resistor modeling for example the
input impedance of  an ultra low noise cryogenic preamplifier. In
this RC circuit, the capacitance corresponds to the capacitive
coupling between the 2DEG and the metallic gate. The voltage across
the resistance ({\it i.e.} before the amplification stage) is the
detected signal unraveling the motion of a single-electron wave
packet beneath the gate. The dissipative element ({\it i.e} the
resistor) is at finite temperature $T$.

An especially important quantity is the signal to noise ratio of the
detector. We derive it by solving for the quantum dynamics of the
circuit when a single electron edge excitation goes through the
gated region. As expected, a high impedance is needed to obtain a
signal to noise ratio of order unity. On average, when the input
impedance is close to the quantum of resistance $R_K=h/e^2$, about
one microwave photon is emitted when a single-electron excitation
moves through the gated region. Since performing quantum operations
on flying qubits requires quantum coherence of the edge excitations,
we have analyzed extensively the backaction of the detector on the
electronic wave packet. The Feynman-Vernon influence functional
\cite{Vernon:1963-1} provides a very convenient tool to discuss
%% Correction syntaxique
damping, spreading and spatial decoherence of the electronic wave packet
induced by the detector. A complete solution is
obtained in the case of chiral edge excitations with linear
dispersion relation. In particular, analytic formulas are obtained
in the low temperature regime where noise in the resistor is
expected to be minimum and dominated by quantum fluctuations.
Contrarily to a continuous weak measurement in which the detector is
constantly interacting with the qubit, the electronic flying qubit interacts
%% correction syntaxique
with our detector during its transit time
through the gated region. Thus, the quantum limit cannot
be defined in terms of the ratio of a measuring time to a
decoherence time. As shown by Clerk {\it et al} \cite{Clerk03},
information theory provides an appropriate framework for defining
the quantum limit. At very low temperatures, a similar analysis is
developed here. An information theoretical measurement efficiency is
defined and expressed in terms of the detector's parameters. It
enables us to find optimal working parameters for the detector so
that information taken away by the detector is most efficiently used
in the detection process. For the present detector, the optimal
measurement efficiency is obtained when the measurement time is not
too large compared with the response time of the detector. A
compromise on the detector's dissipation is needed to optimize the
efficiency while keeping a reasonable signal to noise ratio.

In section \ref{sec:model}, our model for the quantum Hall flying
qubit detector is presented. The signal associated with a
single-electron edge excitation and the corresponding signal to
noise ratio are computed in section \ref{sec:circuit}. In section
\ref{sec:electronic}, we discuss the backaction of the detector on a
single-electron excitation using the Feynman-Vernon influence
formalism. Then, the issue of decoherence is addressed in section
\ref{sec:decoherence}. In section \ref{sec:collision}, we suggest
that a two-electron collision experiment can be used to probe the
spatial decoherence induced by the detector. Consequences for
experiments and conclusions are presented in section
\ref{sec:conclusion}.

%%%%%%%%%%%%%%%%%%%%%%%%%%%%%%%%%%%%%%%%%%%%%%%%%%%%%%%%%%%%%%%%%%%%%%%%%%%%%%%%
\section{Modeling the detector}
\label{sec:model}

\subsection{Presentation of the model}
\label{sec:model:presentation}

\begin{figure}
\includegraphics[width=85mm]{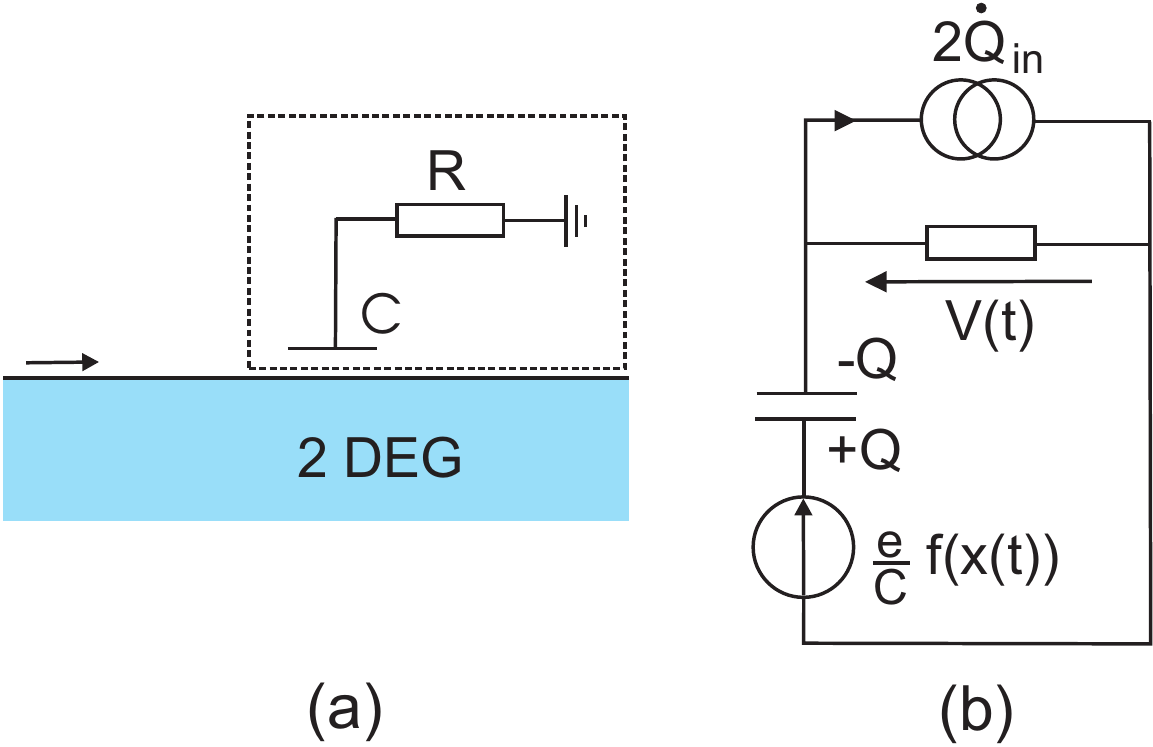}
\caption{\label{fig:circuit} Color online. (a) The detector modeled
as a resistor capacitively coupled to the edge of a 2DEG. (b)
Equivalent circuit for the detector. The edge excitation in the 2DEG
is seen as a transcient voltage source $(e/C)\,f(x(t))$ on the
bottom part of the circuit. Noise in the resistor appears as the
current generator on the top part of the circuit.}
\end{figure}

The detecting device consists into a gate capacitively coupled to
the electronic states of a two dimensional electron gas (2DEG). The
typical gate length  $l$ is of the order of 10 to 100 microns, much
larger than the Fermi wavelength. Any disturbance of the charge
density of the 2DEG induces a voltage across the gate capacitance
which is then amplified. In the present paper, the detector will be
characterized by a resistive input impedance $R$ at effective temperature $T$.
The detection circuit will thus be modeled as an equivalent RC circuit (see Fig.
\ref{fig:circuit}) where $C$ denotes the 2DEG/gate capacitance and
$R$ the input resistance. The detected signal is the voltage drop
$V(t)$ across the resistance. Of course, this
simplified model does not include a detailed description of the
amplifier. But it already contains all the ingredients needed to
discuss the physics of flying qubit detection and in particular, the
backaction of the detector on the qubit. Because we
are studying the single charge detection problem, quantum
fluctuations of the detector might play an important role. Therefore
the detection circuit shall be treated quantum mechanically.

The motion of the electrons in the 2D gas is assumed to be ballistic
and one dimensional. This assumption is satisfied in high mobility
samples where the elastic mean free path may reach the millimeter range.
In the IQHE regime at $\nu=1$, the low energy
excitations of the 2DEG are edge excitations which can be described
as chiral fermions with linear dispersion relation. For the sake of
completeness and having in mind detection of charges within 1D
quantum wires, the case of non relativistic fermions with a
quadratic dispersion relation will also be considered.

Since we are interested by single electron wave packets localized
over a distance $\Delta x$ comparable to the gate size $l$, we will
consider the detection of wave packets injected at energy
$\epsilon_{0}$ above the Fermi energy. The Heisenberg principle
imposes $\epsilon_{0}\gg \hbar v_{F}/l$ where $v_{F}$ is the Fermi
velocity within the considered energy range. Note that in a 2DEG,
the Fermi velocity being of the order of $10^5\
\mathrm{m}\,\mathrm{s}^{-1}$, the typical temperature scale $\hbar
v_{F}/k_{B}l$ is of the order of 100~mK for a 10~$\mu$m gate. In
this regime, the electron is injected far above the Fermi level with
%% correction syntaxique
respect to the energy scales associated with both the temperature and
the detector. Therefore, filled energy levels can be neglected and
the detection problem is reduced to the study of one excess charge
coupled to the quantum RC circuit in a single electron picture.

There are three relevant time scales in the system. The first one is
the circuit response time equal to $RC$. For a high impedance
detector ($R\sim 10^4\ \Omega$) and $C\sim 1$~fF.$\mu m^{-1}$, it is
of the order of 100~ps for a $10~\mu m$ gate and usually scales with
$l$. Having a fast detector implies that this time scale be the
smallest one in the problem. In particular it has to be shorter than
the traveling time $\mathcal{T}=l/v_{F}$ for the excitation below
the gate . This sets the time resolution of the detector. For a 10
to 100~$\mu$m gate, $\mathcal{T}$ is of the order of 100 ~ps to 1~ns
and thus, the fast detection criterion $RC<\mathcal{T}$ is realized
for $R \leq 10^4 ~\Omega$. Finally, the thermal time scale $\hbar
/k_{B}T$ gives the memory time of voltage fluctuations within the RC
circuit at temperature $T$. The $k_{B}T\ll \hbar v_{F}/l$ regime is
called the low temperature regime whereas $k_{B}T\gg \hbar v_{F}/l$
is called the high temperature regime of the detector.

\subsection{Input/output formalism}
\label{sec:model:formalism}

To describe the quantum dynamics of the $RC$ circuit, we follow
the quantum network approach of
Yurke and Denker~\cite{Yurke84} and model the dissipative element
as a semi-infinite ($z \geq 0$) transmission line of characteristic impedance $R$.
Such a transmission line is characterized
by a continuous distribution of capacitance by unit length $C_{T}$
and inductance $\mathcal{L_{T}}$ with specific choice
$\sqrt{\mathcal{L_{T}}/C_{T}} = R$. The transmission line is closed
by the discrete capacitance $C$ at $z=0$. The distributed charge
along the line is denoted by $Q(z,t)$ and describes the transmission
line degrees of freedom. For $z>0$, it satisfies a one-dimensional
wave equation~:
\begin{eqnarray}
\label{eq:circuit:bulk} \mathcal{L_{T}} \frac{\partial ^{2}
Q}{\partial t ^{2}} - \frac{1}{C_{T}} \frac{\partial ^{2}
Q}{\partial z ^{2}} = 0\,.
\end{eqnarray}
The solutions are forward and backward propagating waves with
velocity $v=1/\sqrt{\mathcal{L_T} C_T}$~:
\begin{equation}
 Q(z,t) = Q_{in}(t+ \frac{z}{v}) + Q_{out}(t-\frac{z}{v}),
\end{equation}
whose Fourier decompositions can be written as~:
\begin{eqnarray}
Q_{in}(t) & = & \sqrt{\frac{\hbar}{4\pi R}}\int_{0}^{\infty}
\frac{d\omega }{\sqrt{\omega}} (a_{in}(\omega)e^{-i \omega t
}+\mathrm{h.c.}),\\
Q_{out}(t) & = & \sqrt{\frac{\hbar}{4\pi
R}}\int_{0}^{\infty} \frac{d\omega}{ \sqrt{\omega}}
(a_{out}(\omega)e^{-i \omega t }+\mathrm{h.c.})\,.
\end{eqnarray}
The discrete capacitance $C$ appears as a boundary condition on
$z=0$ for the transmission line~:
\begin{equation}
\frac{1}{C_{T}}\frac{\partial Q}{\partial z}(0,t) -\frac{Q(0,t)}{C}
 =0\,.
\end{equation}
Using the notation $Q(t)$ for the charge of the capacitor at the end of the line
$Q(t)=Q(z=0,t)$, the boundary condition can be rewritten as~:
\begin{equation}
\label{eq:circuit:boundary}
 R \dot{Q}(t) + \frac{Q(t)}{C} = 2
R\dot{Q}_{in}(t)\,.
\end{equation}
An electron propagating in the 2DEG underneath the gate induces a
charge $q_{ind}$ on the gate. In a single particle approach, it can
be written as $q_{ind}(t) = e f(x(t))$, where $x(t)$ is the position
of the electron with respect to the gate at time $t$. The function
$f$ accounts for the shape of the gate.
%%%%%%%%%%%%%%%%%%%%%%%%%%%%%%%%%%%%%%%%%%%%%%%%%%%%%%%%%%%%%%%%%%
A realistic device will have a typical gate size of the order of
10~$\mu$m to be compared to a spacing of 100~nm between the 2DEG and
the gate. Thus the 2DEG-gate coupling in the neighborhood of $x=0$
will be close to unity~: almost all the electric field lines will go
from the passing electron to the gate. Thus we will only consider
the simplified case of optimal coupling $f(x=0) = 1$ (we note that
the less than optimal situation leads to similar results). Far away
from the detector, $f(x \rightarrow \pm \infty ) = 0$ as the
electron does not feel the presence of the gate.
%%%%%%%%%%%%%%%%%%%%%%%%%%%%%%%%%%%%%%%%%%%%%%%%%%%%%%%%%%%%%%%%%%
The typical width of the function $f(x)$ is the size of the the gate
$l$, much greater than the Fermi wavelength. In full generality
$f(x)=h(2x/l)$ where $h(u)$ decays rapidly for $|u|\leq 1$. In this
paper, a triangular shape function will often be used to obtain
explicit results. It is defined by $f(x)=0$ for $|x|\geq l/2$ and
$f(x)=1-2|x|/l$ for $|x|\leq l/2$. Although the shape function $f$
is not universal, this specific case will enable us to get explicit
analytical results which capture the essential physics of the
problem.

\medskip

The induced charge appears as a source term added to the noise term in
\eqref{eq:circuit:boundary}:
\begin{equation}
\label{eq:circuit:boundary:total}
R \dot{Q}(t) + \frac{Q(t)}{C} = 2 R\dot{Q}_{in}(t) +
\frac{e}{C}f(x(t))\,.
\end{equation}
The equations of motion
 \eqref{eq:circuit:bulk} and \eqref{eq:circuit:boundary:total} can be derived
from a Lagrangian of the form:
\begin{equation}
\label{eq:model:Lagrangian}
L= L_{el}  + L_{0}+ L_{\partial}
\end{equation}
where $L_{el}$ is the Lagrangian of the electronic system
and $L_{0}$ is the bulk Lagrangian for the
transmission line:
\begin{equation}
L_{0} = \int_{0}^{+\infty}\left[ \frac{\mathcal{L_{T}}}{2}\left(
\frac{\partial Q }{\partial t}\right)^{2} - \frac{1}{2C_{T}} \left(
\frac{\partial Q }{\partial z}\right)^{2}\right]\,dz
\end{equation}
The boundary Lagrangian $L_{\partial}$
corresponds to the electrostatic energy stored in the capacitance and describes the
interaction between the charge $Q(0,t)$ and the induced charge:
\begin{equation}
L_{\partial} = - \frac{1}{2C}\,\Big[ Q(0,t)-ef(x(t)) \Big]^2\,.
\label{Full}
\end{equation}
The RC circuit degrees of freedom are the modes propagating along the
transmission line. They are easily quantized by imposing the usual
commutation relations for the bosonic modes~:
\begin{eqnarray}
[{a}_{\alpha}(\omega),{a}^{+}_{\beta}(\omega')] =
\delta(\omega-\omega') \delta_{\alpha,\beta},
\end{eqnarray}
where $\alpha$,$\beta$ stand for $in$ or $out$ operators. The Lagrangian
\eqref{eq:model:Lagrangian} will be used within a path integral formalism when discussing
decoherence of the electronic wavepacket.

\medskip

In this paper, two possible forms for the electronic
Lagrangian $L_{el}$ will be considered: {\it (i)}  the case of a chiral electron with relativistic
(linear) dispersion appropriate for describing the edge excitations of a $\nu =1$ droplet
and {\it (ii)} the case of a non relativistic particle with parabolic dispersion. The latter will allow a
more complete understanding of the specifics of chiral edge excitations.

%%%%%%%%%%%%%%%%%%%%%%%%%%%%%%%%%%%%%%%%%%%%%%%%%%%%%%%%%%%%%%%%%%%%%%%%%%%%%%%%
\section{Evolution of the circuit}
\label{sec:circuit}

In this section, the evolution of the detector during the measurement process is discussed.
As a first step, we shall consider a classical charge moving along a fixed trajectory in the 2DEG.
The signal generated by such a moving charge will be computed and expressed in term of
a number of microwave photons sent into the detector. This signal will be compared to
the input noise generated by the preamplifier. We will provide explicit
expressions for the corresponding signal to noise ratio.

\subsection{Detected signal}
\label{sec:circuit:signal}

\subsubsection{Quantum signal in the input/output formalism}
\label{sec:circuit:input-output}

Let us compute the circuit's evolution when a charge $e$ travels in the 2DEG
along a given trajectory $t\mapsto x(t)$. Solving for the time evolution
equation \eqref{eq:circuit:boundary:total}
gives access to the voltage drop in the
transmission line $V(t)=-(1/C_{T})\,\, \partial  {Q}/\partial z=(e f(t) -Q(t))/C$. This
voltage is the input signal for the preamplifier and is equal to~:
\begin{eqnarray}
V(t) & = & -\frac{1}{C}\sqrt{\frac{\hbar}{\pi R}} \int_{0}^{\infty}
\frac{\sqrt{\omega}}{\omega -1/(iRC)} \,\,  {a}_{in}(\omega)
 \,\, e^{-i\omega t} +\mathrm{h.c} \nonumber\\
 & + & \frac{e}{RC^{2}} \int_{0}^{\infty} d\tau
 \,\, e^{-\frac{\tau}{RC}}  \,\, \big[f(t)-f(t-\tau)\big],
\label{V}
\end{eqnarray}
where $f(t)$ stands for $f(x(t))$.
The first term corresponds to the quantum noise generated by the resistor and
the second term is the signal created by the external charge.
The average voltage is obtained by tracing over the
transmission line degrees of freedom. Only the second term of Eq.(\ref{V}) contributes to
the average voltage~:
\begin{eqnarray}
\label{eq:circuit:average-voltage}
\langle  {V}(t)\rangle =  \frac{e}{RC^{2}} \int_{0}^{\infty}
d\tau  \,\, e^{-\frac{\tau}{RC}} \,\, \big[f(t)-f(t-\tau)\big] .
\end{eqnarray}
A good detector is expected to react
faster than the characteristic time $\mathcal{T}=l/v_F$ of the
electron-gate interaction. Assuming that
the electron velocity is very close to the Fermi velocity $v_{F}$, the fast detector condition
is  $RC<\mathcal{T}$.
For a fast detector, the average detected voltage becomes $\langle  V(t) \rangle = R e\,\frac{df}{dt}$.

\subsubsection{Photon emission}
\label{sec:circuit:photons}

%% Correction
A rough estimate for photon emission into the detection circuit
can be obtained noticing that the typical
voltage generated by a single charge is $eR/\mathcal{T}$. This electromagnetic pulse lasts during time
$\mathcal{T}$. Thus, the total energy emitted is of the order
$R^{-1}(eR/\mathcal{T})^2\mathcal{T}\sim e^2R/\mathcal{T}$. Assuming photons
have a frequency $\mathcal{T}^{-1}$, the typical number
of photons emitted into the detector during the measurement is
$\bar{n}_{\mathrm{m}}\sim
e^2R/\hbar=2\pi R/R_K$ where $R_K=h/e^{2}$ is the resistance quantum.
A first implication of this result is the need for
a high impedance detector in order to limit the total acquisition time.

Let us denote by $\mathcal{E}(\omega))$ the spectral density of the
energy dissipated in the resistor during measurement of a single
charge traveling at fixed velocity $v_{F}$. It is given in terms of the Fourier
transform $\tilde{V}(\omega)$ of the average
voltage $\langle V(t)\rangle $ by $\mathcal{E}(\omega)
=|\tilde{V}(\omega)|^2/2\pi R$. Then, Eq.
\eqref{eq:circuit:average-voltage} leads to:
\begin{equation}
\label{eq:photons:density}
\mathcal{E}(\omega)=\hbar\,\frac{R}{R_K}\,
\frac{\omega^2}{\omega^2+\frac{1}{(RC)^2}}\,
\left(\frac{l}{2v_{F}RC}\right)^2\,\left| \tilde{h}
\left(\frac{\omega l}{2v_{F}}\right)\right|^2\,.
\end{equation}
where the gate shape function $f$ is written as $f(x)=h(2x/l)$ and
$\tilde{h}$ denotes the Fourier transform of $h$. This expression
shows that the RC circuit acts as a high pass filter cutting
frequencies below $1/RC$. The shape of the gate acts as a low pass
filter. For a fast detector, the latter dominates and photon
emission preferably takes place around frequency $v_{F}/l$.

\subsection{Signal to noise ratio}
\label{sec:circuit:SNR}

\subsubsection{General expressions}

The signal to noise ratio is an important characteristic of a detector. In the present case,
a first estimate can already be obtained using the above photon number estimations.
In the low temperature limit $k_{B}T\ll \hbar v_{F}/l$, most of the modes where
photon emission take place are unpopulated. Therefore,
we expect $\bar{n}_{\mathrm{m}}$ to provide a crude estimate for the signal to noise ratio of the measurement
device which would then be of the order of $R/R_K$.
In the high temperature regime $k_{B}T\gg \hbar v_{F}/l$, all the relevant modes will be populated
by $l\,k_{B}T/\hbar v_{F}$ photons corresponding to the thermal
noise of the detector. Therefore, the signal to noise ratio should reduced by
this factor in the high temperature regime. This discussion also shows that the
low temperature regime is also a quantum regime with respect to the
modes relevant for detection.

Let us now turn towards a more precise definition of the signal to noise
ratio. Let consider as our signal the voltage collected over $\mathcal{T}/2$~:
\begin{equation}
\label{eq:SNR:signal}
\overline{V}(t)=\frac{2}{\mathcal{T}}\int_{t}^{t+\mathcal{T}/2}d\tau
 {V}(\tau)\,.
\end{equation}
Contrarily to the voltage collected over a time $\mathcal{T}$, this quantity
does not vanish on average.
Assuming a triangular gate function of width $l$ and a fast detector, the maximum average signal when the
electron passes through the gate is given by~:
\begin{equation}
\langle \overline{V}\rangle = \frac{2\,R\,e}{\mathcal{T}}
\int_{0}^{\mathcal{T}/2}d\tau  \frac{df}{d\tau}
 =  \frac{2Re}{\mathcal{T}} .
\end{equation}
where bracketing $\langle \ldotp \rangle$ denotes quantum
statistical averaging at temperature $T$. The fluctuation
$\Delta\overline{V}^2=\langle \overline{V}^2\rangle -\langle
\overline{V}\rangle^2$ is directly related to the symmetrized noise
correlator $g(\tau) = \langle \lbrace V_{I}(\tau), V_{I}(0)
\rbrace_{+}\rangle/2$ where $V_{I}$ denotes the voltage operator in
the interaction representation. Explicitly, it is given by~:
\begin{equation}
\label{g}
g(\tau) = \frac{\hbar}{\pi RC^{2}} \int_{0}^{\infty}
\coth(\frac{\hbar
\omega}{2k_{B}T}) \,
\frac{\omega\,\cos(\omega \tau)}{\omega ^{2}+ (\frac{1}{RC})^{2}}\,d\omega\,.
\end{equation}
which, for a triangular gate of width $l$, leads to~:
\begin{equation}
\label{eq:SNR:noise-integral} \Delta\overline{V}^{2} =
\frac{\hbar}{\pi RC^{2}}\int_{0}^{\infty}
\coth(\frac{\hbar\omega}{2k_{B}T})\, \frac{\sin^2{\left(\frac{\omega
\mathcal{T}}{4}\right)}}{(\frac{\omega \mathcal{T}}{4})^{2}}\,
\frac{\omega\,d\omega}{\omega ^{2}+ (\frac{1}{RC})^{2}} .
\end{equation}
Finally, the signal to noise ratio is defined as~:
\begin{equation}
\mathrm{SNR}_{T}=\frac{\langle \overline{V}\rangle^2}
{\langle \overline{V}^2\rangle -\langle \overline{V}\rangle^2}
\end{equation}

\subsubsection{Asymptotics and numerical results}

At vanishing temperatures and for a fast detector, the signal to noise ratio saturates to a value given by~:
\begin{equation}
\label{eq:SNR:zero-T} \mathrm{SNR}_{0} \simeq \frac{R}{R_K}\;
\frac{\pi^{2}}{\log \big[\frac{l}{2v_FRC}\big] + \gamma},
\end{equation}
where $\gamma$ is the
Euler constant $\gamma \approx 0.577$. The logarithmic correction with respect
to $2\pi R/R_K$ shows that the signal to noise ratio is optimized by decreasing
the measurement time within the limits of fast measurement.

\begin{figure}
\includegraphics[width=85mm]{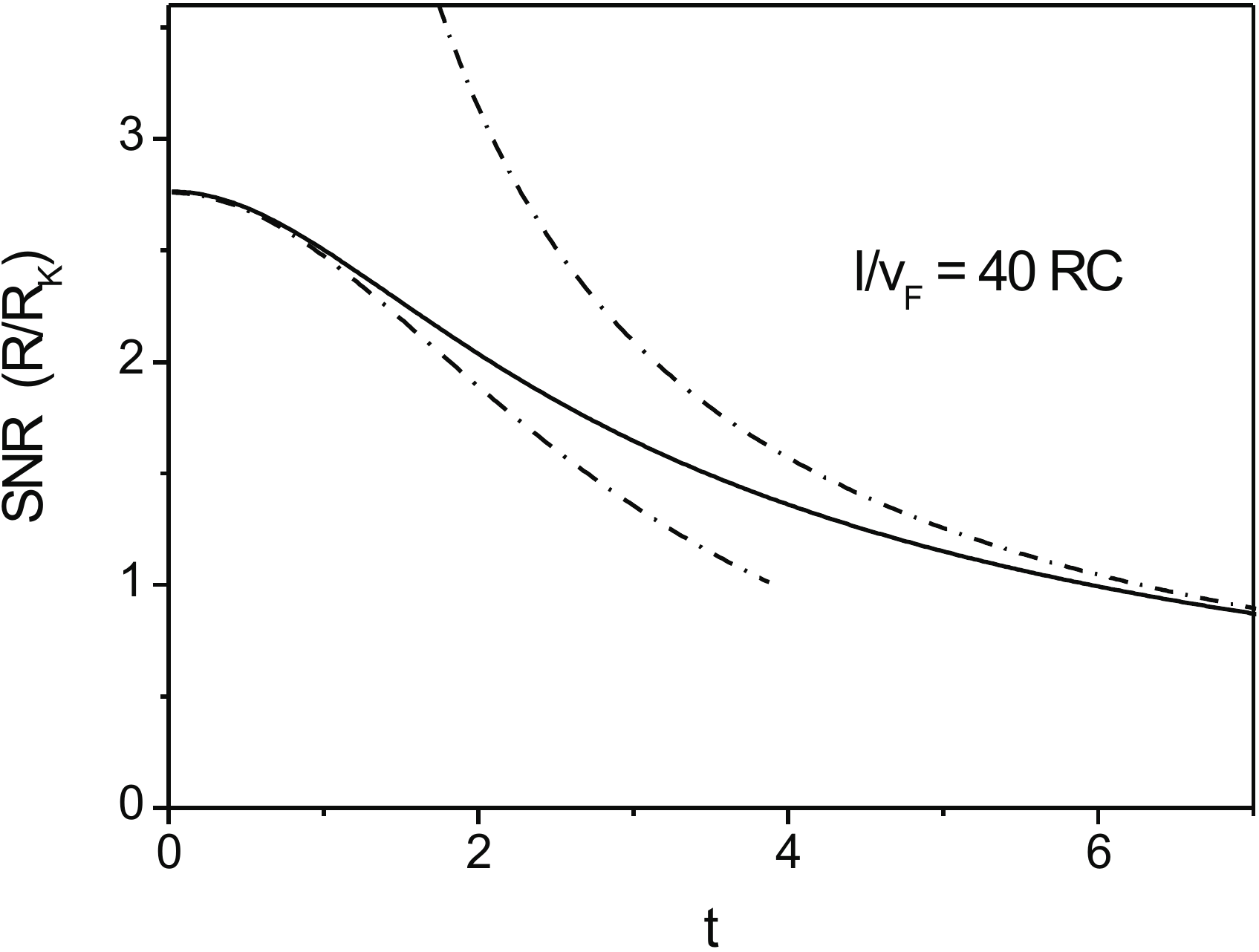}
\caption{The signal to noise ratio divided by $R/R_K$
as a function of the dimensionless temperature $\mathrm{t}k_{B}T/(\hbar v_{F}/l)$ and for $\l/v_{F}=40\,RC$ (full lines). Dashed dotted lines represent the high and
low temperature asymptotics respectively given by
Eq. \eqref{eq:SNR:high-T} and Eqs. \eqref{eq:SNR:zero-T} and \eqref{eq:SNR:low-T:basic}.}
\label{fig:1}
\end{figure}

As expected and as shown on Fig.~\ref{fig:1}, the signal to noise
ratio decreases with increasing temperature. In the very low
temperature regime $k_{B}T\ll \hbar v_F/l \ll\hbar/RC$, the
temperature dependence can be extracted by noticing that, in the
noise integral \eqref{eq:SNR:noise-integral}, $\coth{(\hbar
\omega/2k_{B}T)}-1$ departs from zero in a frequency range much
below $v_{F}/l$ and $1/RC$. We obtain~:
\begin{equation}
\label{eq:SNR:low-T:basic}
\frac{1}{\mathrm{SNR}}_{T}-\frac{1}{\mathrm{SNR}}_{0} =
\frac{R_K}{24\,R}\,\left(\frac{k_{B}T}{\hbar v_{F}/l}\right)^2\,.
\end{equation}
In the high temperature limit $k_{B}T>>\hbar v_F/l$, we have~:
\begin{equation}
\label{eq:SNR:high-T} \mathrm{SNR}_{T}\simeq \; 2 \pi \;
\frac{R}{R_K} \; \frac{\hbar v_F/l}{k_{B}T }.
\end{equation}
Having characterized the evolution of the detection circuit as well as its signal to noise ratio, we will now
turn to its backaction on the electronic degrees of freedom.

%%%%%%%%%%%%%%%%%%%%%%%%%%%%%%%%%%%%%%%%%%%%%%%%%%%%%%%%%%%%%%%%%%%%%%%%%%%%%%%%
\section{Evolution of the electronic degrees of freedom}
\label{sec:electronic}

In this section, the quantum backaction of the detector on the
edge excitations will be analyzed. Within a path integral
formalism, the Feynman-Vernon approach~\cite{Vernon:1963-1} gives access to
the electronic evolution by integrating out the RC circuit degrees of
freedom through the so-called influence
functional. This functional contains all information
on dissipation (energy relaxation) and decoherence
caused by the interaction with the detector.
After a brief reminder of the Feynman-Vernon formalism, the non relativistic (quadratic dispersion relation)
and chiral fermion (linear dispersion relation) cases will be successively considered.

\subsection{General formalism}

The Feynman-Vernon formalism is based on an explicit
integration over the environmental degrees of freedom
treated as an effective bath of harmonic oscillators in
thermal equilibrium. Let us denote the initial electronic reduced density operator by $\rho_{0,el}$
and $\rho_{0,r}$ the one associated with the detector.  Then, at time $t\geq 0$, the electron reduced
density operator is given by:
\begin{equation}
\rho_{el}(x^+_{f},x^-_{f},t) = \langle
x^+_{f}|\,\mathrm{Tr}_{r}\left( U(t)\,(
\rho_{0,el}\otimes\rho_{0,r})\, U^{+}(t)\right) \,
|x^-_{f}\rangle\,. \label{evolro}
\end{equation}
Note that here, a factorized initial condition is appropriate since the electron injection is
performed on demand at a given time \cite{Feve07}.
In full generality, this reduced density operator will evolve according to:
\begin{equation}
\label{eq:electron:kernel}
\rho_{el} (x^+_{f},x^-_{f},t)  =  \int
J_{t}\left(
\begin{array}{c}
x^+_{f}\\
x^-_{f}
\end{array}
\begin{array}{c}
x^+_{i}\\
x^-_{i}
\end{array}
\right)
\,\rho_{0,el}(x^+_{i},x^-_{i})\,dx^+_{i} dx^-_{i}
\end{equation}
The evolution kernel can be expressed within a path integral formalism
in terms of the action for a free electron moving in the classical potential created by
the gate $S_{el}=\int
(L_{el}-e^{2}f(x)^2/2C) d\tau$ and in terms of the Feynman-Vernon influence functional
$\exp{(-\Phi_{FV}[x^+,x^-])}$ representing the backaction of the detector~:
\begin{equation}
J_{t}\left(
\begin{array}{c}
x^+_{f}\\
x^-_{f}
\end{array}
\begin{array}{c}
x^+_{i}\\
x^-_{i}
\end{array}
\right)
  =  \int\mathcal{D}[x^+,x^-]\,
e^{\frac{i}{\hbar}S_{el}[x^+,x^-]}\, e^{-\Phi_{FV}[x^+,x^-]}\,.
\label{FFV}
\end{equation}
where $S_{el}[x^+,x^-]=S_{el}[x^+]-S_{el}[x^-]$.
In this equation, the boundary conditions are given by
$x^\pm(t)=x^\pm_{f}$ and $x^\pm(0)=x^\pm_{i}$.
Considering the resistor as in equilibrium with temperature $T$,
the Feynman-Vernon influence functional
$\exp{(-\Phi_{FV}[x^+,x^-])}$ is Gaussian~:
\begin{widetext}
\begin{eqnarray}
\Phi_{FV}[x^+,x^-] & = & \frac{e^{2}}{\hbar RC^2} \int_{0}^{t} d\tau
\int_{0}^{\tau}du \Big[f[x^+(\tau)]-f[x^-(\tau)]\Big]
\Big[L(\tau-u)f[x^+(u)] -\bar{L}(\tau-u)f[x^-(u)]\Big],
\label{PhiFV}
\\
L(\tau-u) & =& \int_0^{\infty} \frac{d\omega}{\pi} \frac{\omega}{\omega^{2}+
(1/RC)^{2}} [\coth(\frac{ \hbar \omega}{2k_{B}T})\cos(\omega
(\tau-u)) -i \sin(\omega (\tau-u))].
\label{bigL}
\end{eqnarray}
\end{widetext}
The real part of $\Phi_{FV}$ is responsible for decoherence and the
imaginary part describes dissipation. The $\coth(\frac{ \hbar \omega}{2k_{B}T})$
factor that relates them is a direct consequence of the detailed balance condition
expressing that the gate is initially at equilibrium with temperature $T$.
Note that the free evolution action $S_{el}[x^+,x^-]$ depends whether chiral
fermions in the Quantum Hall Regime or non relativistic fermions with quadratic dispersion relation
are considered.

\subsection{Quadratic dispersion relation}
\label{sec:electron:quadratic}

We will now study the usual case of free fermions with a quadratic
dispersion relation $\epsilon_{k}=\hbar^2 k^{2}/2m$ where $m$ is the electron
effective mass. In the high
temperature regime $k_{B}T\gg\hbar v_{F}/l$, the problem is greatly simplified
since the Feynman-Vernon influence functional becomes local in time.
The spreading of the electronic wavepacket and the energy damping
caused by the gate can be calculated in a semiclassical approach.

\medskip

As shown in appendix \ref{appendix:Langevin}
the probability distribution for the particle is then described by a
Langevin equation that can be directly derived from the Feynman-Vernon formalism:
\begin{eqnarray}
\label{eq:quadratic:Langevin}
m\ddot{r} +  e^{2}R f'(r)^{2}\dot{r} &= & f'(r) \xi(t)\,.
\end{eqnarray}
where $\xi(t)$ denotes a classical Gaussian noise:
\begin{eqnarray}
\label{eq:quadratic:Langevin:noise}
\langle \xi(t_{1}) \xi(t_{2}) \rangle & =& 2R e^{2}k_{B}T
\delta(t_{1}-t_{2})
\end{eqnarray}

\subsubsection{Energy relaxation}
\label{sec:electron:quadratic:damping}

Eq. \eqref{eq:quadratic:Langevin} shows that interaction with the gate also introduces a position dependent
friction force proportional to the electronic velocity. This friction force leads to
the energy relaxation for the particle. Assuming weak dissipation, the velocity will
remain close to $v_{F}$ and thus, approximating
$f'(r)\approx f'(v_Ft)$ leads to an exponential damping of the velocity~:
\begin{equation}
\langle v(t)\rangle = v_F \; e^{-\frac{e^{2}R}{m}\int_{0}^{t} d\tau
f'(v_F\tau)^{2}}
\end{equation}
In the case of a triangular gate of width $l$, the above expression gives the velocity drop after
the particle has crossed the gate~:
\begin{equation}
\frac{\langle v_{\mathrm{after}}\rangle}{\langle v_{\mathrm{before}}
\rangle} =  \exp{\left(-4\frac{R}{R_K}\frac{\lambda_{F}}{l}\right)},
\end{equation}
where  $\lambda_{F}=\hbar / mv_{F}$ denotes the Fermi wavelength
(typically $\lambda_{F}\sim 20$~nm in AsGa). Since $l\gg
\lambda_{F}$, the weak dissipation assumption is satisfied for
$R\lesssim R_K$.

\subsubsection{Spreading of the wave packet}
\label{sec:electron:quadratic:spreading}

Let us now focus on the evolution of the width  $\delta r^{2} =
\langle(r-\langle r\rangle)^{2}\rangle$ of the electronic wavepacket
due to the gate. Neglecting the electronic velocity damping, the
spreading of the wave packet $\Delta_{\delta r^2}(t)=\delta r^{2}(t)
-  \delta r^{2}(0)$ is given by~:
\begin{equation}
\Delta_{\delta r^2} (t)= \frac{4Re^{2}k_{B}T}{
m^{2}}\int_{0}^{t}
d\tau_{1}\int_{0}^{\tau_{1}}d\tau_{2}\int_{0}^{\tau_{2}}d\tau_{3}
f'(v_F\tau_{3})^{2}\,.
\end{equation}
In the case of a triangular gate, the total spreading
$\Delta_{\delta r^2} $ due to the interaction with the gate is given
by~:
\begin{eqnarray}
\Delta_{\delta r^2} = \frac{8}{3}\frac{R}{R_K}\frac{ k_BT}{hv_{F}/l}
\lambda_{F}^{2}\,.
\end{eqnarray}
Scaling like $\lambda_{F}^{2}$, the wavepacket spreading is expected to be small although it might be
observable for high enough temperature.

The full evolution of the electron reduced density matrix can be described in
the high temperature regime using a Markovian master equation.
Contrarily to the usual quantum Brownian motion, decoherence and frictions
terms depend on the gate shape function $f$ making an exact solution
impossible. An approximate solution to
this equation describing decoherence induced by the gate at high temperatures is
given in appendix \ref{appendix:master-equation}. However, it is not valid in the low
temperature regime potentially relevant for future experiments. As we shall see
now, in the case of chiral fermions with linear dispersion relation, an exact solution
can easily be obtained leading to a fully general expression for the spatial decoherence induced
by the measurement.

\subsection{Chiral fermions}
\label{sec:decoherence:chiral}

\subsubsection{Exact real time evolution}

Let us now consider chiral fermions with linear
dispersion. In this case, the static potential $e^{2}f(x)^{2}/2C$
created by the gate does not lead to any backscattering but only to an
additional forward scattering phase with respect to the free chiral fermion evolution~:
\begin{equation}
U_{0}(t) |x \rangle = e^{-\frac{i\,e^{2}}{2\hbar C
v_F}\int_{x}^{x+v_Ft}du f(u)^{2}}\, |x +v_Ft \rangle .
\end{equation}
Therefore, the position
operator evolves according to the free ballistic evolution for chiral fermions propagating at
the Fermi velocity $v_{F}$: $x(t)= {x}(0)+v_Ft$.
Because of this very simple evolution of the position
operator, the r.h.s. of eq. \eqref{evolro} can be evaluated by going to the interaction scheme with respect
to the gate/electron interaction. Introducing $x_{i}^\pm=x_{f}^\pm-v_{F}t$, we have~:
\begin{equation}
\rho_{el}(x^+_{f},x^-_{f},t) = \rho_{0,el}(x^+_{i},x^-_{i})\,
e^{i(\phi_{+}-\phi_{-})}\,\mathcal{D}_{t}(x^+_{f},x^-_{f})\,.
\label{eq:evolution:chiral}
\end{equation}
where the phase $e^{i(\phi_{+}-\phi_{-})}$ is given by
$\phi_{\pm}=\frac{e^2}{2\hbar
C}\int_{0}^tf(x_{f}^\pm-v_{F}\tau)^2\,d\tau$ and corresponds to the
forward scattering phase induced by the gate. Because of chirality
and linear dispersion for the edge excitations, dissipation only
introduces a multiplicative decoherence factor
$\mathcal{D}_{t}(x^+_{f},x^-_{f})$ given by~:
\begin{equation}
\mathcal{D}_{t}(x^+_{f},x^-_{f}) = \mathrm{Tr}_r\left(
U_{I}[x^+_{f},t]\,\rho_{0,r}\, U^{\dagger}_{I}[x^-_{f},t]\, \right)
\label{rhochir}
\end{equation}
where $U_{I}[x,t] ={T} e^{-i\frac{e}{\hbar C}\int_{0}^{t}
f(x-v_F(t-\tau)) {Q}_{I}(\tau)d\tau}$.
Eq. \eqref{rhochir} shows that this decoherence coefficient is simply
given by Feynman-Vernon influence functional evaluated for the classical
trajectories $x_{\pm}(\tau)=x^\pm_{i}+v_{F}\tau$. Therefore we have~:
\begin{equation}
\mathcal{D}_{t}(x^+_{f},x^-_{f}) =  e^{-\Phi_{FV}[x^+_{f},x^-_{f}]},
\label{eq:FV:chiral:0}
\end{equation}
where
\begin{widetext}
\begin{eqnarray}
\Phi_{FV}[x^+_{f},x^-_{f}] & = & \frac{e^{2}}{\hbar RC^2} \int_{0}^{t}
d\tau \int_{0}^{\tau}d\tau' \Big[f[x^+_{f},\tau]-f[x^-_{f},\tau]\Big]
\Big[L(\tau-\tau')f[x^+_{f},\tau'] -\bar{L}(\tau-\tau')f[x^-_{f},\tau']\Big],
\label{eq:FV:chiral:1}\\
f[x,\tau] & =& f\big[x-v_F(t-\tau)\big].
\label{eq:FV:chiral:2}
\end{eqnarray}
\end{widetext}
Equations \eqref{eq:evolution:chiral} and \eqref{eq:FV:chiral:0} to
\eqref{eq:FV:chiral:2} provide the complete solution for the
dynamics of a chiral edge excitation with exact linear dispersion in
the presence of the detector modeled as an RC circuit. Remember that
our "on the fly" detector influences the chiral edge excitation only
while traveling near the gate. Because of this finite interaction
time, $x_{f}^\pm$ are chosen on the right side of the gate,
corresponding to positions after interaction with the detector.
Accordingly, the time scale $t$ is chosen so that
$x_{i}^\pm=x_f^\pm-v_{F}t$ be on the left side of the date. This is
equivalent to extending integration limits in eq.
\eqref{eq:FV:chiral:1} to the half plane $\tau'\leq \tau$.

\subsubsection{Dissipation}

Before considering how the detector affects the spatial
quantum coherence of the chiral fermion, it is important to discuss energy dissipation associated with
the detection process. As explained in section \ref{sec:model:presentation}, the effect of the Fermi sea
has been neglected because the extra charge has been injected high enough above the Fermi level.
It is thus important to check that this assumption remains valid through the detection process.
The discussions of section \ref{sec:circuit:photons} and \ref{sec:electron:quadratic:damping}
already show that this is indeed the case.
Focusing on the low
temperature regime which gives the best signal to noise ratio and assuming $R\lesssim R_K$,
the typical energy emitted is of the order of $(R/R_K)\times (\hbar v_{F}/l)\lesssim
\hbar v_{F}/l$. This corresponds to the energy
loss by the detected electron and therefore, since $\epsilon_{0}\gg \hbar v_{F}/l$, the
detection process does
not bring it back to the Fermi level.

Finally, the coupling to the gate leads to a destruction of
spatial coherence existing prior to the interaction with the gate.
A detailed study of the spatial decoherence induced by the gate is presented in
section \ref{sec:decoherence} and its consequences on a two electron collision experiment
are presented in section \ref{sec:collision}.

%%%%%%%%%%%%%%%%%%%%%%%%%%%%%%%%%%%%%%%%%%%%%%%%%%%%%%%%%%%%%%%%%%%%%%%%%
\section{Decoherence of the electronic wavepacket}
\label{sec:decoherence}
%%%%%%%%%%%%%%%%%%%%%%%%%%%%%%%%%%%%%%%%%%%%%%%%%%%%%%%%%%%%%%%%%%%%%%%%%

\subsection{General discussion}
\label{sec:decoherence:general}

In this section, we will focus on the decay of spatial coherence
for the chiral edge excitation. It is directly related to the real
part of the Feynman-Vernon exponent $\Phi_{FV}(x^+_{f},x^-_{f})$ ~:
\begin{equation}
\left|
\frac{\rho_{el}(x^+_{f},x^-_{f},t)}{\rho_{0,el}(x^+_{i},x^-_{i})}
\right|
=e^{-\Re\big[\Phi_{FV}(x^+_{f},x^-_{f})\big]}
\label{eq:chiral:decoherence}
\end{equation}
Since the coupling to the detector takes place during a short window
of time corresponding to the passing time $l/v_{F}$ of edge
excitations beneath the gate, equation \eqref{eq:chiral:decoherence}
will be used te evaluate the decay of spatial coherence after the
whole detection process. This is achieved by taking a wide enough
range of integration in the explicit expression for
$\Phi_{FV}(x^+_{f},x^-_{f})$ given by Eqs. \eqref{eq:FV:chiral:1}
and \eqref{eq:FV:chiral:2}. The resulting decoherence exponent
$\Gamma_{c}(d)=\Re{(\Phi_{FV}(x^+_{f},x^-_{f}))}$ then only depends
on the distance $d=|x^+_{f}-x^-_{f}|$ over which spatial coherence
is probed and on the detector's parameters (size $l$, response time
$RC$ and temperature $T$). Its general expression is~:
\begin{equation}
\Gamma_{c}(d)=\frac{e^{2}}{\hbar RC^2}
\int_{0\leq \tau'\leq \tau}\Delta_{f}(\tau)\,
\Re{(L(\tau-\tau'))}\Delta_{f}(\tau')\,d\tau\,d\tau'
\label{eq:decoherence:Gamma}
\end{equation}
where $\Delta_{f}(\tau)=f[x^+_{f},\tau]-f[x^-_{f},\tau]$. In the
large separation limit $d\rightarrow \infty$, only diagonal
contributions associated with products $f[x,\tau]\,f[x,\tau']$
survive and give the residual coherence~:
\begin{equation}
\label{eq:decoherence:general:d-infinite} \Gamma_{c}(\infty)=
\frac{2\,e^{2}}{\hbar RC^2} \int_{\tau'\leq \tau} f[x,\tau]\,
\Re{(L(\tau-u))}\,f[x,\tau']\,d\tau \,d\tau'
\end{equation}
which is indeed independent of $x$. Note that this contribution only
probes the noise kernel of the detector over the detection time
scale $l/v_{F}$.
When $d$ decreases, the off diagonal contribution involving products
$f[x_{f}^\pm,\tau]\,f[x_{f}^\mp,\tau]$ in \eqref{eq:decoherence:Gamma}
become important. It is sensitive to the noise over a time scale $d/v_{F}$. Assuming that
the noise evaluated at $|\tau-\tau'|=d/v_{F}$ varies smoothly over $l/v_{F}$, an estimate
for the off diagonal contribution is then~:
\begin{equation}
\label{eq:decoherence:general:large-d}
\Gamma_c^{(\mathrm{od})}(d)\simeq -\frac{\pi\eta}{2}\frac{R}{R_K}
\,\left(\frac{l}{RCv_{F}}\right)^2\,
\Re{\left(
 L\left(\frac{d}{v_F}\right)
\right)}
\end{equation}
where $\eta=(\int h(u)\,du)^2$ accounts for the precise shape of the gate. Assuming the RC circuit
has memory time $\tau_{m}$, saturation of decoherence will be actually reached for $d\gtrsim v_{F}\tau_{m}$.

For small separation $d\ll l$, the decoherence exponent is expected to scale as $(d/l)^2$.
The exact expression can be derived be expanding
$\Delta(\tau)=\frac{d}{v_{F}}\,\frac{\partial f}{\partial \tau}[\bar{x}_{f}]$ where
$x_{f}^\pm=\bar{x}_{f}\pm d/2$:
\begin{equation}
\label{eq:decoherence:general:small-d} \Gamma_{c}(d)
=\frac{e^{2}d^{2}}{\hbar RC^2 v_{F}^{2}} \int_{\tau'\leq \tau} \! \!
\! \! \! \! \! \! \! d\tau\,d\tau'\,
\frac{df}{d\tau}[\bar{x}_{f},\tau]\,\Re{(L(\tau-\tau'))}\,
\frac{df}{du}[\bar{x}_{f},\tau'] .
\end{equation}
The decoherence exponent will now be evaluated and discussed in the various temperature
regimes of the detector. Asymptotic expressions will be derived in the case of $d\rightarrow \infty$
and $d\ll l$ and compared to numerical evaluations of $\Gamma_{c}(d)$ for the case of a triangular gate.

\subsection{Results for a triangular gate}

\begin{figure}
\includegraphics[width=90mm]{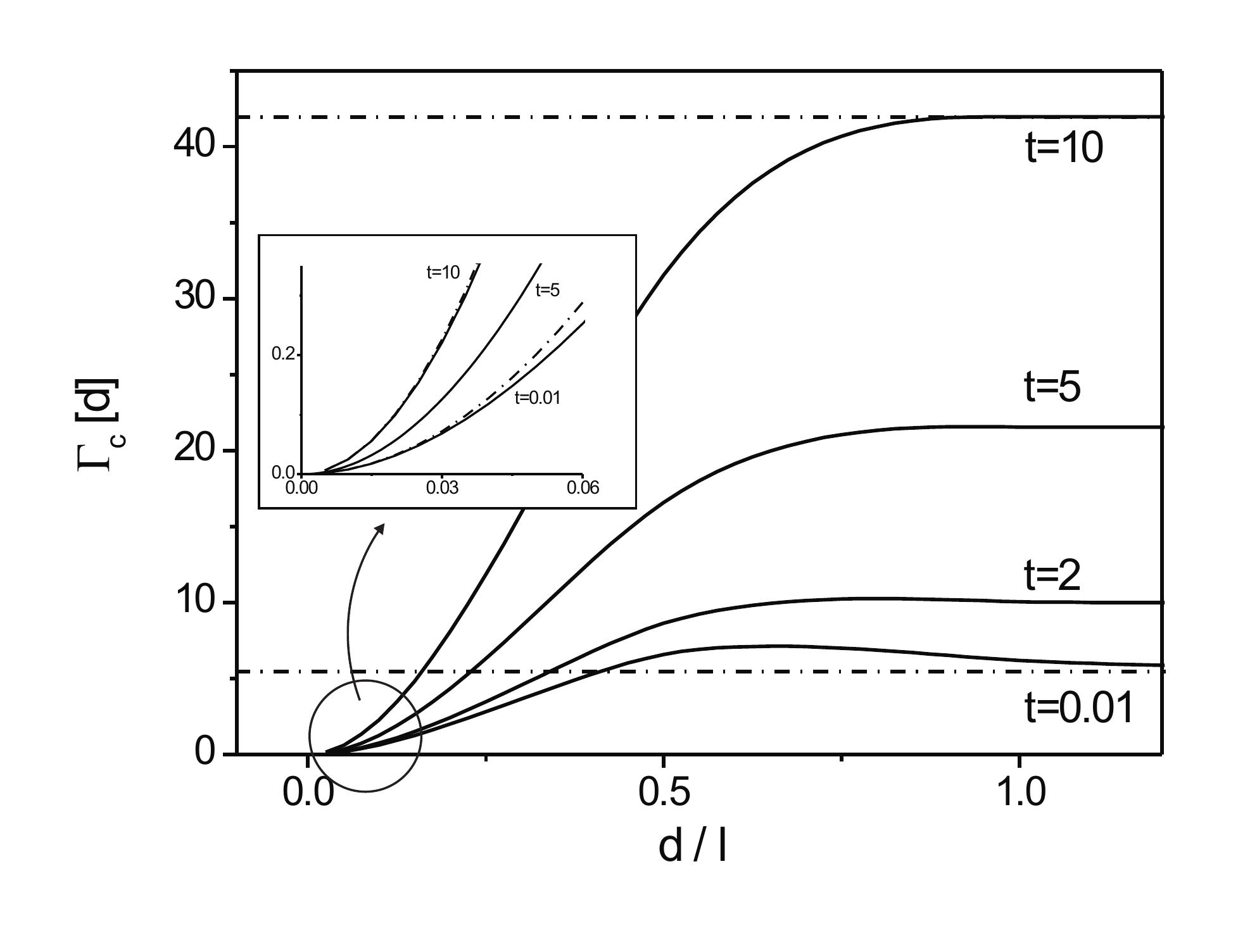}
\caption{\label{fig:2} Decoherence exponent $\Gamma_{c}(d)$ as a function of $d/l$
for various dimensionless temperatures $t=k_{B}T/(\hbar v_{F}/l)$ for $l/v_{F}=40\,RC$ (full line).
The inset shows the behavior at short distance with tyical $(d/l)^2$ behavior. Dotted lines show
the $d\ll l$ and $d\gg l$ asymptotics given by Eqs. \eqref{eq:decoherence:zero-T:small-d}
and \eqref{eq:decoherence:zero-T:infinity} for very low temperature and
by Eqs. \eqref{eq:decoherence:high-T:small-d} and \eqref{eq:decoherence:high-T:infinity}
}
\end{figure}

In the case of a triangular gate, an integral expression valid in all regimes is easily obtained~:
\begin{equation}
\Gamma_{c}(d) = 16 \frac{R}{R_K} \int_0^{+\infty} \frac{
\sin^4{(\lambda_{l} x)}\,\sin^2{(2\lambda_{d}x)} }
{\lambda_{l}^2 x^4
}
\, \frac{x\,\coth{\left(\alpha\,x\right)}}{x^{2}+1}
\,dx\,.
\end{equation}
where $\lambda_l =l/4v_{F}RC$ and $\alpha= \hbar(RC)^{-1}/2k_{B}T$.
This expression is plotted on Fig. \ref{fig:2} as a function of
$d/l$ for various temperatures.

\subsubsection{Zero temperature limit}
\label{sec:decoherence:low_T}

For small separation $d\ll l$, we have
\begin{equation}
\label{eq:decoherence:zero-T:small-d} \Gamma_c (d) \simeq 24 \big[
\log( \frac{l}{2 v_F RC}) + \gamma -\frac{\log(2)}{3} \big]
\;\frac{R}{R_K} \; \left(\frac{d}{l}\right)^2\,.
\end{equation}

For large separation $d\gg l$, the decoherence exponent saturates at
\begin{equation}
\label{eq:decoherence:zero-T:infinity} \Gamma_c (\infty) = 8 \log(2)
\; \frac{R}{R_K}\,.
\end{equation}
As mentioned above, the memory time of the RC circuit is responsible
for the large $d/l$ behavior of the decoherence coefficient (see eq.
\eqref{eq:decoherence:general:large-d}). At low temperature, the
noise kernel of the $RC$ circuit behaves as $-1/\pi \times
(RC/(\tau-\tau'))^2$ and therefore this shows that decreasing $d$
leads to an increase of the decoherence exponent:
\begin{equation}
\Gamma_c^{(\mathrm{od})}(d)\simeq \frac{\eta}{2}\,\frac{R}{R_K}
\left(\frac{l}{d}\right)^2\,.
\end{equation}
Equivalently, it means that decreasing $d$ from a very large value first leads to a stronger
decoherence.
%% Correction
This result is a direct consequence of the long range correlations present
in the $RC$ circuit at very low temperatures. This effect is expected to be weak ($l/d\ll 1$)
%% Correction syntaxique
and disappears rapidly with increasing temperature.

\subsubsection{High temperature limit}
\label{sec:decoherence:high_T}

In the high temperature limit, the noise kernel can be considered as
local in time $\Re{(L(s))}\simeq 2(RC)^2\,(k_{B}T/\hbar)\,\delta(s)$
and therefore the decoherence exponent is explicitly given by
\begin{eqnarray}
\label{eq:decoherence:high-T} \Gamma_{c}(d) & = &
2\pi\frac{R}{R_K}\frac{k_{B}T}{\hbar v_{F}/l}\,\int
\Delta(y,d/l)^2\,\frac{dy}{l}\\
\Delta(y,d/l) & = & h\left(y+\frac{d}{l}\right)-h\left(y-\frac{d}{l}\right)\,.
\label{eq:decoherence:high-T:aux}
\end{eqnarray}
Note that the coefficient in front of geometrical factor $\int
\Delta(y,d/l)^2\,dy/l$ scales as the thermal photon number
$\frac{k_{B}T}{\hbar v_{F}/l}$ reflecting the increase of
fluctuations with temperature. As expected, the decoherence exponent
saturates at $d\rightarrow \infty$ and also scales as $(d/l)^2$ for
$d\ll l$. In the case of a triangular gate function, these
expressions can be explicitly evaluated in the small $d\ll l$ and
large distance $d\gg l$ limits:
\begin{eqnarray}
\label{eq:decoherence:high-T:small-d}
\Gamma_{c}(d \rightarrow 0) & \simeq &  8 \pi \; \frac{R}{R_K} \; \frac{k_BT}{\hbar v_{F}/l} \;
\left(\frac{d}{l}\right)^2\\
\Gamma_{c}(\infty) & = &\frac{4 \pi}{3} \; \frac{R}{R_K} \;
\frac{k_BT}{\hbar v_{F}/l} \,.
\label{eq:decoherence:high-T:infinity}
\end{eqnarray}

\subsection{Efficiency and quantum limit for flying qubit detection}

We are now in position to characterize the efficiency of the flying
qubit detection. But in the present detection scheme, the electronic
edge excitation and the detector interact during a finite time
$l/v_{F}$. Therefore, the notion of quantum limit cannot be defined
by referring to the measurement and dephasing rates which do not make
sense here. Indeed, as stressed in the introduction, it should be
approached through the basic concepts of information theory. Clerk
{\it et al} have shown that the quantum limit can be understood from
an information theory perspective \cite{Clerk03}: the quantum limit
is reached when the information extracted by the detector is
entirely used for the measurement.

We will now elaborate on this general idea having in mind the optimization
of flying qubit detection.
%% Correction syntaxique
For simplicity, the discussion will
focus on the case of chiral edge excitations with linear dispersion
relation. We will first discuss
the accessible information stored in the detector while probing two edge excitations initially
located at two different positions. Then,
the ability of the measurement scheme to distinguish between the two initial positions using
the voltage signal \eqref{V} will be discussed quantitatively using information theory.
The detector efficiency will then be defined in terms of these two notions and will be used to discuss
its optimization as well as its ability to reach the quantum limit.

\subsubsection{Accessible information}

Let us consider the interaction between a coherent wave packet described by
the wavefunction $\psi_{0}$ at time $t_{i}=0$ and the detector.
At zero temperature, the electron + detector state at time $t$ is precisely of the
form generalizing Eq. (14) of \cite{Clerk03}:
\begin{equation}
\int \psi_{0}(x)\,|x+v_{F}t\rangle \otimes |D_{t}(x)\rangle\,dx
\end{equation}
where
\begin{equation}
|D_{t}(x)\rangle = Te^{-\frac{i}{\hbar}\int_{0}^t
f(x+v_{F}\tau)\,Q_{I}(\tau)\,d\tau}\,|0\rangle\,.
\end{equation}
As seen above, since $f(x+v_{F}\tau)=0$ at large enough times, the
$t$ dependence can safely be dropped out. Each state $|D(x)\rangle$
is a tensor product of coherent states over all the modes
propagating in the transmission line $|D(x)\rangle\otimes_{\omega}|\alpha_{\omega}(x)\rangle$. The information
relative to the position $x$ of the edge excitation is stored in the
phases of the complex amplitudes characterizing these coherent
states.

Different initial positions $x^+$ and $x^-$ of the edge excitation
correspond to different phases and this leads to the measurement induced decoherence
$e^{-\Gamma_{c}(d)}=|\langle D(x^-)|D(x^+) \rangle|$
where $d=|x^+-x^-|$. This decoherence coefficient
is indeed an infinite product of contributions corresponding to all
the propagating modes of the transmission line:
\begin{equation}
\label{eq:decoherence:spectral-decomposition}
e^{-\Gamma_{c}(d)}=\left|\prod_{\omega=0}^\infty
\langle \alpha_{\omega}(x^-)|\alpha_{\omega}(x^+)\rangle\,\right|\,.
\end{equation}
Therefore, introducing the spectral density $\Gamma_{c}(d,\omega)$ of the decoherence
exponent $\Gamma_{c}(d)$,
the contribution of scalar products associated with modes
within $[\omega,\omega+d\omega]$ in the r.h.s. of Eq.
\eqref{eq:decoherence:spectral-decomposition} is equal to:
\begin{equation}
\prod_{\omega\leq \omega'\leq\omega+d\omega}^\infty
|\langle \alpha_{\omega'}(x^-)|
\alpha_{\omega}(x^+)\rangle|\sim 1-\Gamma_{c}(d,\omega)\,d\omega\,.
\end{equation}
Therefore, summing
Eq. (24) of Clerk {\it et al} \cite{Clerk03} over all the modes,
the quantum mechanical accessible information associated with the pair of states
$|D(x^+)\rangle$ and $|D(x^-)\rangle$ is given by:
\begin{equation}
\mathcal{I}[x^+,x^-]=\int_{0}^\infty
\Gamma_{c}(d,\omega)\,d\omega=\Gamma_{c}(d)\,.
\end{equation}
This equation shows that the decoherence exponent has a direct
information-theoretic interpretation.

\subsubsection{Measurement information}

Let us now provide an estimate for the information associated
with the measurement process. For the sake of simplicity, we will restrict
ourselves the problem of distinguishing between two initial
positions $x^+$ and $x^-$ separated by a distance $d$. These
corresponds to the two possible inputs of the communication channel
associated with the complete detection device.

Detection of an edge excitation is performed through the classical signal
average over a time $\mathcal{T}/2$ (see Eq. \eqref{eq:SNR:signal}). In the case of
a fast detector $RC\ll \mathcal{T}$, the signal
created by an edge excitation initially located at $x$ is given by:
\begin{equation}
\label{eq:decoherence:signal}
\overline{V}(t)=\frac{2eRv_{F}}{l}
\left[
f(x+\frac{l}{2}+v_{F}t)-f(x+v_{F}t)
\right]\,.
\end{equation}
Let us now consider two average signals $\overline{V}_{\pm}(t)$
associated with initial positions $x^\pm$ separated by a distance
$d$. The difference between these two signals can be used to
distinguish between $x^+$ and $x^-$. It clearly follows from
\eqref{eq:decoherence:signal} that the maximum difference at a given
time $t$ varies from $(4Rev_{F}/l)\times (d/l)\times
\mathrm{max}(|h'(u+l/2)-h'(u)|)$ for $d\ll l$ to $4Rev_{F}/l$ for
$d=l/2$ and then decreases to $2eRv_{F}/l$ when $d\gg l$.

Because of the noise
within the detector, the signal
at time $t$ is distributed according to a Gaussian whose variance is given by
\eqref{eq:SNR:noise-integral}. Therefore, the mutual information associated with two signals having
$|\overline{V}_{+}-\overline{V}_{-}|
=\mathrm{max}_{t}(|\overline{V}_{+}(t)-\overline{V}_{-}(t)|)$ and variance $\Delta\overline{V}$
can be expressed as
\begin{eqnarray}
R(\lambda) & = & \log{(2)}
-  \frac{\lambda}{2}\int_{0}^\infty
e^{-\frac{(\lambda z)^2}{8}}
F(\lambda,z)
\,\frac{dz}{\sqrt{2\pi}} \\
F(\lambda,z) & = &
\log{ \left((1+e^{-\frac{\lambda^2}{2}(z+1)})(1+e^{\frac{\lambda^2}{2}(z-1)})\right)}
\end{eqnarray}
where $\lambda=|\overline{V}_{+}-\overline{V}_{-}|
/\Delta \overline{V}$.
%% Correction syntaxique
In general, $\lambda$ depends on $d$, the details of the gate shape function and the signal to noise
ratio computed in section \ref{sec:circuit:SNR}.
The measurement information $\Gamma_{m}(d)=R(\lambda(d))$
represents the maximum rate of information
collected through the detector if one modulates the edge excitation sources by sending trains of
either delayed or advanced localized electrons separated by a time $d/v_{F}$.

\medskip

For $d\ll l$, up to some coefficient that reflects the details of
the gate shape function, $\lambda \simeq
4\sqrt{\mathrm{SNR}_{0}}\times (d/l)$. In the opposite limit $d\gg
l$, $\lambda \sim \sqrt{\mathrm{SNR}_{0}}$. The behavior of
$R(\lambda)$ in various limits is also very simple. For $\lambda
\rightarrow \infty$, $R(\lambda)\rightarrow \log{(2)}$. It reflects
the fact that the measurement delivers exactly one bit of
information since it enables to distinguish unambiguously between
$x^+$ and $x^-$. In the $\lambda \ll 1$ limit, $R(\lambda)\simeq
\lambda^2/8$ which leads to a $(d/l)^2$ scaling for $d\ll l$: the
detector hardly distinguishes two very close initial positions.

\subsubsection{Measurement efficiency and optimization}

Knowing the accessible information
$\Gamma_{c}(d)$ and the measurement information
$\Gamma_{m}(d)$,
we define the measurement efficiency as $\Gamma_{m}(d)/\Gamma_{c}(d)$.

\medskip

At short distances ($d\ll l/\sqrt{\mathrm{SNR}_{0}}$), we have
\begin{equation}
\Gamma_{m}(d)\sim 2\,\mathrm{SNR}_{0}\times (d/l)^2
\end{equation}
and therefore the
efficiency of the measurement becomes independent from $R/R_K$ and
only depends on the size of the gate since $\Gamma_c(d)$
scales as $(R/R_K)\,(d/l)^2$.
Using expression \eqref{eq:SNR:zero-T} for the zero temperature signal to
noise ratio, this leads to:
\begin{equation}
\frac{\Gamma_{m}(d)}{\Gamma_{c}(d)}\simeq \frac{\pi^2/12\,}
{\left(\gamma-\frac{\log{(2)}}{3}+\log{\left(\frac{l}{2v_{F}RC}\right)}\right)\,
\left(\gamma+\log{(\frac{l}{2v_{F}RC})}\right)}\,.
\end{equation}
Optimizing the detector requires the choice of a low value for
$l/v_{F}RC$ and in this case, this also corresponds to the
optimization of the signal to noise ratio for a fixed $R/R_K$.

\medskip

Contrarily to the short distance case, the efficiency at large
distances ($d\gg l$) depends on $R/R_K$. At very large $R/R_K$, the
signal to noise ratio increases indefinitely but this does not
improve the measurement information. The high decoherence then leads
to a low efficiency that goes as $R_K/8R$. More generally, we have
shown that $\mathrm{SNR}_{0}=(R/R_K)\times
\kappa_{\mathrm{snr}}(l/v_{F}RC)$ where $\kappa_{\mathrm{snr}}$
depends on the shape of the gate. In the same way, for $d\rightarrow
\infty$, Eq. (\ref{eq:decoherence:general:d-infinite}) tells us that
at zero temperature, $\Gamma_{c}(\infty)= (R/R_K)\times
\kappa_{c}(l/v_{F}RC)$. Thus, the measurement efficiency at large
distances is given by~:
\begin{equation}
\label{eq:decoherence:efficiency:d-infinity}
\frac{\Gamma_{m}(\infty)}{\Gamma_{c}(\infty)} =
\frac{\kappa_{\mathrm{snr}}}{8\,\kappa_c} \times
\frac{8\,R(\sqrt{\mathrm{SNR}_0})}{\mathrm{SNR}_0}\,.
\end{equation}
Remarkably, the ratio of
\eqref{eq:decoherence:efficiency:d-infinity} to its very weak
coupling value is a universal function of the signal to noise
ratio~:
\begin{equation}
\frac{(\Gamma_{m}/\Gamma_{c})(\infty)}{
\lim\limits_{R/R_K\rightarrow 0}((\Gamma_{m}/\Gamma_{c})(\infty))}=
\frac{8\,R(\sqrt{\mathrm{SNR}_{0}})}{\mathrm{SNR}_{0}}\,.
\end{equation}

\begin{figure}
\includegraphics[width=85mm]{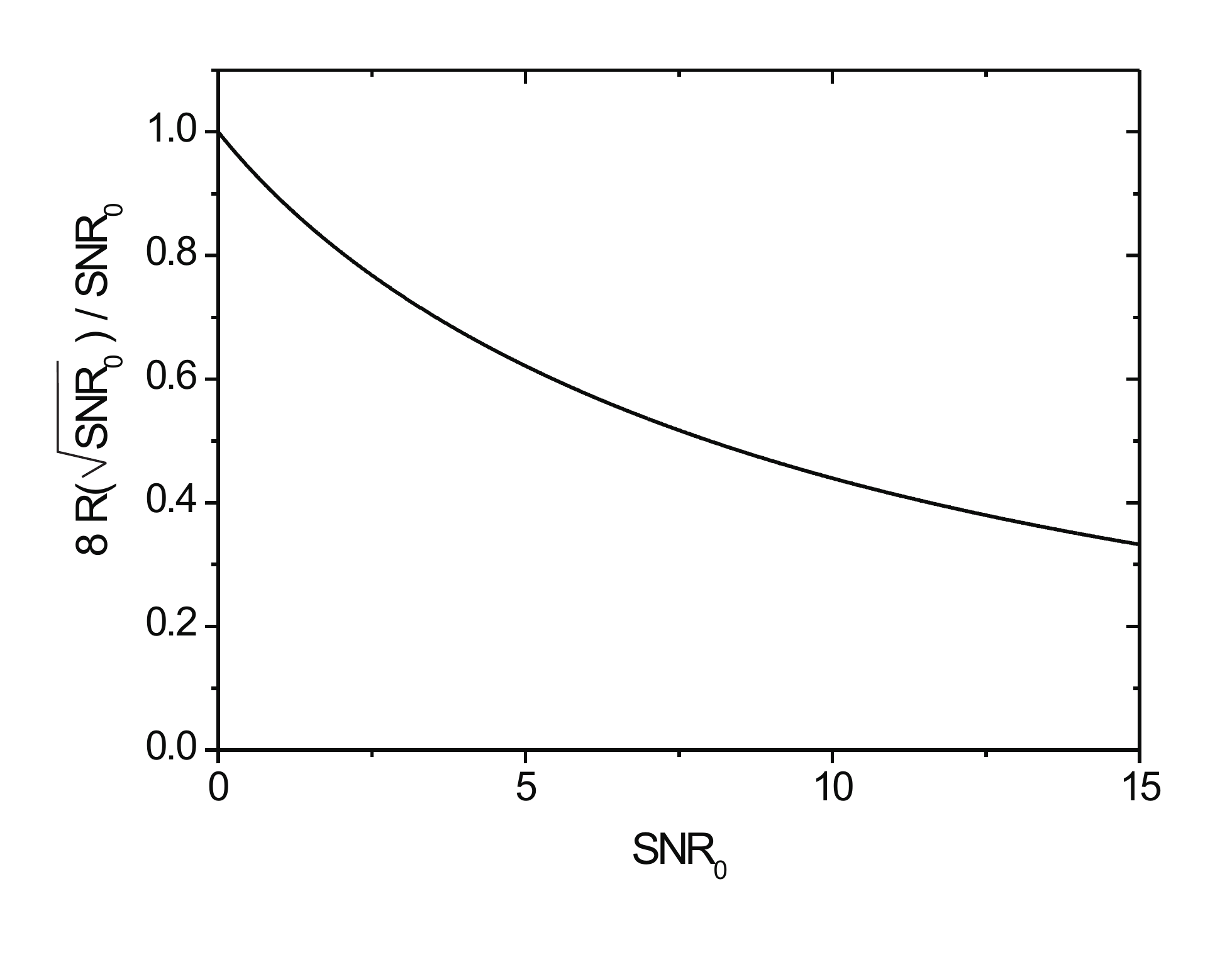}
\caption{\label{fig:eff-snr}Ratio of the large distances ($d
\rightarrow \infty$) efficiency for arbitrary coupling $R/R_K$ to
its value in the very weak coupling limit ($R/R_K<<1$). }
\end{figure}
This function is plotted on Fig. \ref{fig:eff-snr}. It suggests that
a reasonable signal to noise ratio $\mathrm{SNR}_{0}\sim 1$ can be
reached without loosing too much in terms of efficiency compared to
the weak coupling situation. In the limit of a weakly coupled
detector ($R\ll R_K$), the efficiency tends to
$\kappa_{\mathrm{snr}}/8\kappa_c$ which depends on the gate shape.
It is generically a decreasing function of $l/2v_{F}RC$. As in the
short distance case, decreasing $l/v_{F}RC$ at fixed $R/R_K$
corresponds to optimizing the mode structure in order to minimize
decoherence while ensuring an efficient measurement.

\medskip

\begin{figure}
\includegraphics[width=85mm]{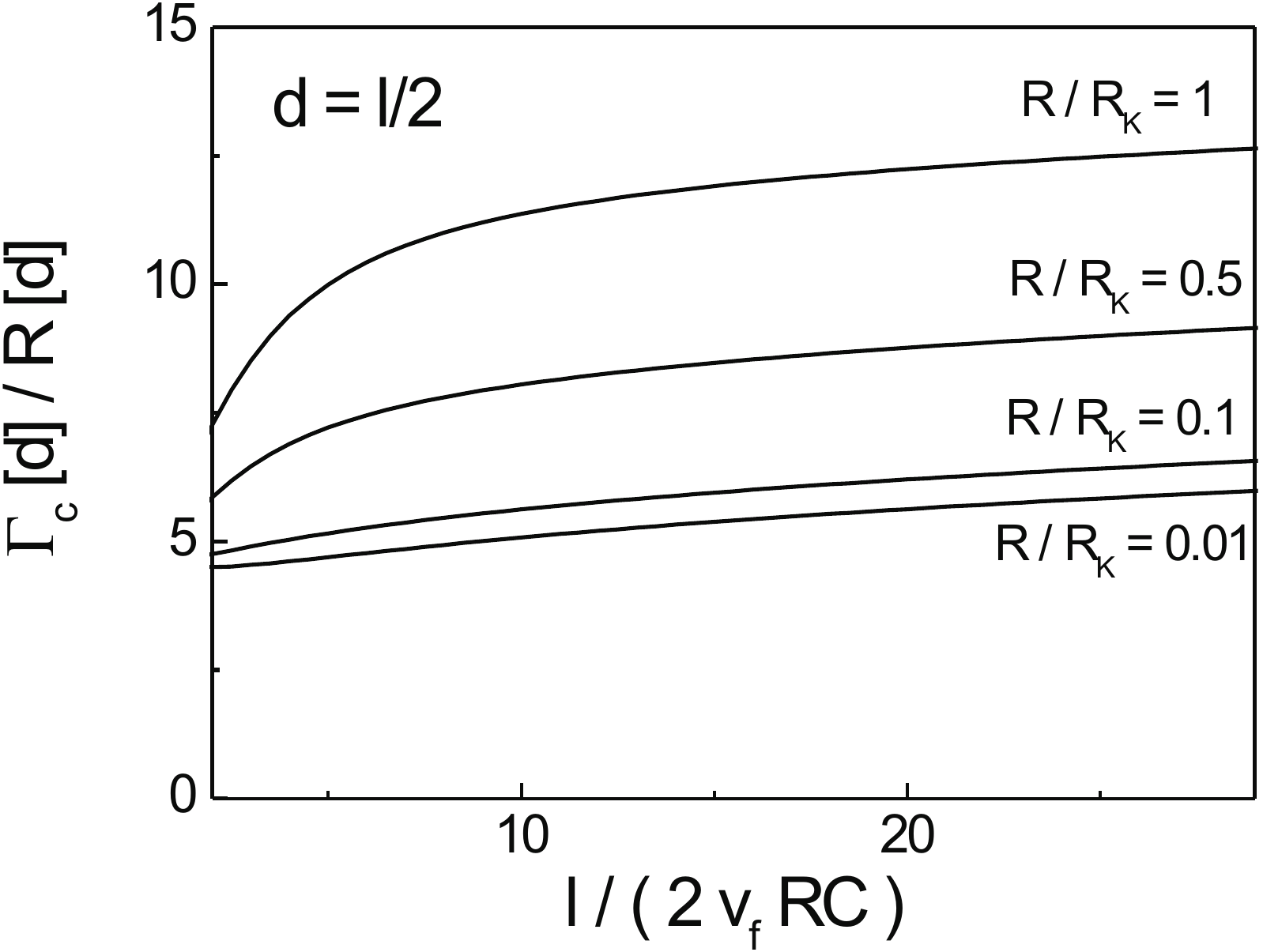}
\caption{\label{fig:efficiency-dl2} Inverse of the efficiency as a function of $l/2v_{F}RC$
for various values of $R/R_K$. These curves are computed for a triangular gate
taking into account the finiteness of $RC$ and not making the fast detector approximation.}
\end{figure}

Fig. \ref{fig:efficiency-dl2} shows the
inverse of the efficiency computed for $d=l/2$ as a function of $l/2v_{F}RC$
for a triangular gate, taking into account the exact formula \eqref{eq:circuit:average-voltage}
integrated over time $\mathcal{T}/2$ so that the finite response time of the circuit is taken into
account.

These curves exhibit a rather weak dependence in $l/2v_{F}RC$ even
down to values where the fast detector approximation is not expected
to work. They also show that going from a weakly coupled detector
$R/R_K\lesssim 0.1$ to a high impedance one $R/R_K\sim 1$ for
$l/2v_{F}RC\sim 5$ leads to a drop of the efficiency by 50~\% while
increasing the signal to noise ratio by a factor 10. Typical values
range from $0.2$ to $0.1$ which shows that the quantum RC circuit
does not reach the quantum limit in the sense of
$\Gamma_{m}(l/2)/\Gamma_{c}(l/2)\sim 1$. A slower detector would
reach an efficiency of $0.25$ in the weak coupling regime and $0.15$
at $R/R_K=1$.

%%%%%%%%%%%%%%%%%%%%%%%%%%%%%%%%%%%%%%%%%%%%%%%%%%%%%%%%%%%%%%%%%%%%%%%%%%%%%
\section{Effect of decoherence on a two electron collision}
\label{sec:collision}
%%%%%%%%%%%%%%%%%%%%%%%%%%%%%%%%%%%%%%%%%%%%%%%%%%%%%%%%%%%%%%%%%%%%%%%%%%%%%

We now describe an experiment which enables detecting decoherence of
the electronic wavepacket by the detector. The idea is to perform a
two electron collision through a quantum point contact that acts as
a beam splitter for the electronic edge excitations. In the case of
bosons, photons bunch together and tend to come out through the same
quantum channel. Only recently has bunching of photons emitted by
independent sources been experimentally demonstrated by Ph.~Grangier
and his collaborators \cite{Beugnon:2006-1}. Here, because of the
Pauli principle, antibunching is expected: the electrons should
leave into different outgoing channels after their passing through
the beam splitter. On the other hand, classical particles should
partition randomly thus leading to a probability of $1/4$ for
leaving both in a given outgoing channel, and $1/2$ for leaving in
different channels. As we shall see, introducing the flying qubit
%% Correction syntaxique
detector on one of the input channels leads to an intermediate
behavior interpolating between full antibunching and the partition
behavior depending on the decoherence introduced by the detector.
We consider only the case of complete spin polarization as appropriate
for a 2DEG in the IQHE regime.
%% Correction syntaxique
The realization of an electronic Mach-Zehnder interferometer
operating precisely in these conditions has recently been
performed~\cite{Heiblum07}.

\begin{figure}
\includegraphics[width=85mm]{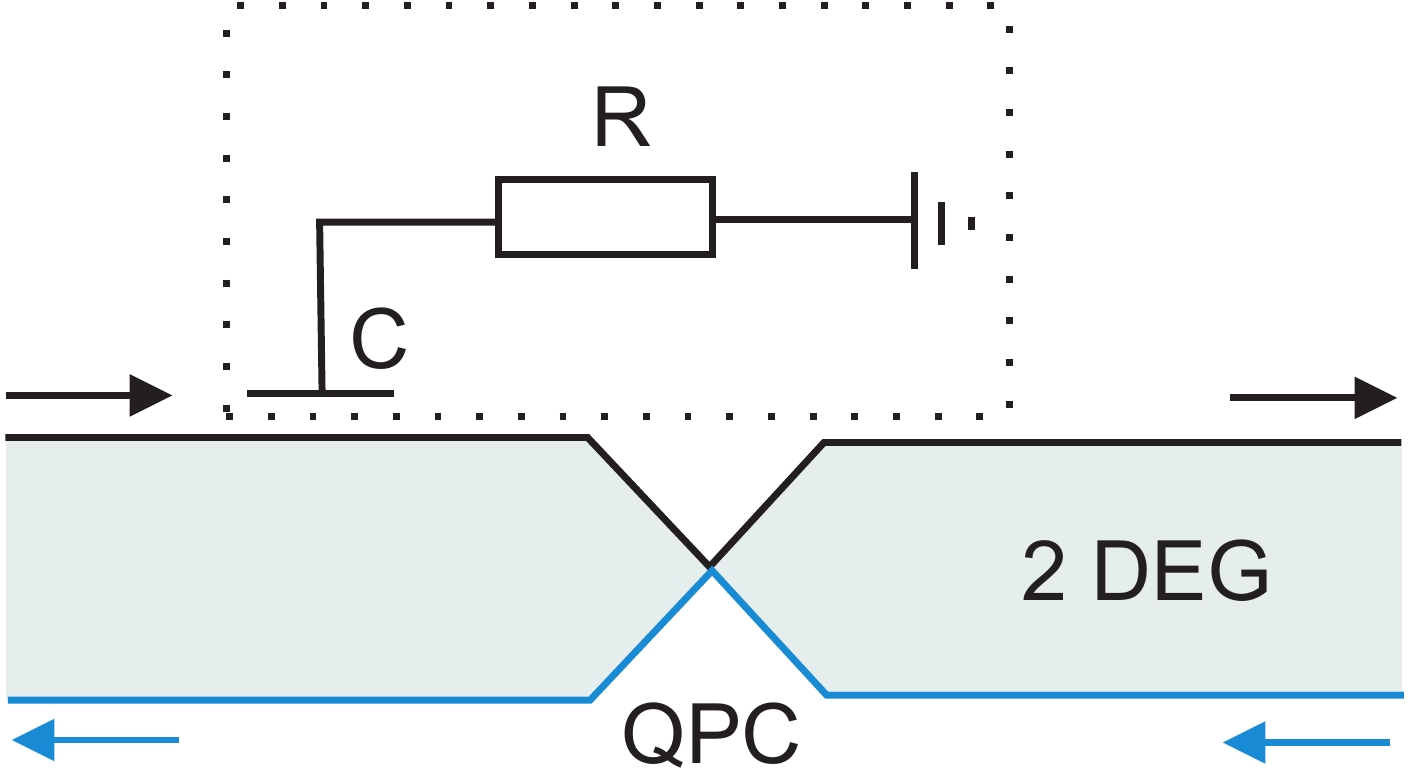}
\caption{\label{fig:two-excitations} Color online. Principle scheme
for a two electron collision experiment within a 2DEG. Electronic
edge excitations are injected into the incoming edge channels
(incoming arrows) and detected on the outgoing channels (outcoming
arrows) after having passed through the QPC. A detector (dotted box)
is located on one of the incoming edge channels.}
\end{figure}

The geometry of the experiment is described on fig.
\ref{fig:two-excitations}~: two edge states labeled by $+$ and $-$
correspond to the different directions of propagation (tracks) on
the opposite sides of the sample are connected through a quantum
point contact (QPC) located at $x=0$. Such a quantum point contact
tuned at transmission $1/2$ realizes the analog of an optical beam
splitter. Denoting the state corresponding to the position $x$ on
track $\pm$ by $|x,\pm\rangle$, the QPC action on one particle
states is given by:
\begin{eqnarray}
|x_{i},+\rangle
& \rightarrow &\frac{1}{\sqrt{2}}(|x_{f},+\rangle+|x_{f},-\rangle) \\
|x_{i},-\rangle
& \rightarrow & \frac{1}{\sqrt{2}}(|x_{f},+\rangle-|x_{f},-\rangle)
\end{eqnarray}
assuming that $x_{i}$ is located before the QPC with respect to electron propagation
and $x_{f}=x_{i}+v_{F}t$ is
located after it. The flying qubit detector is located on one of the input channels which are
fed with the same wavefunction $\psi(x+x_{0})$ centered around a position $x_{i}=-x_{0}$
where
\begin{equation}
\psi(x) = \left(\frac{2}{\pi a^{2}}\right)^{1/4}
e^{ik_Fx}e^{-\frac{x^{2}}{a^{2}}} \,.
\label{wfunc}
\end{equation}

Antibunching of the particles is probed through the probabilities of joint detection
after collision on the beam splitter on the same outgoing
edge channel (let say $+$), $P_{++}$ or on opposite edge channels
$P_{+-}$.
The probability $P_{++}$ can be computed as the probability to find
both electrons on edge state $+$ at position $x_{f}-\delta/2\leq x \leq
x_{f}+\delta/2$ where $x_{f}$ is the central position of the detecting
area and $\delta \gg a$ so that this detecting area fully overlaps the
electronic wavepackets. For the probability $P_{+-}$, indiscernability
of the particles is taken into account by symmetrizing the
projection operator under the exchange of both particles.

\begin{figure}
\includegraphics[width=85mm]{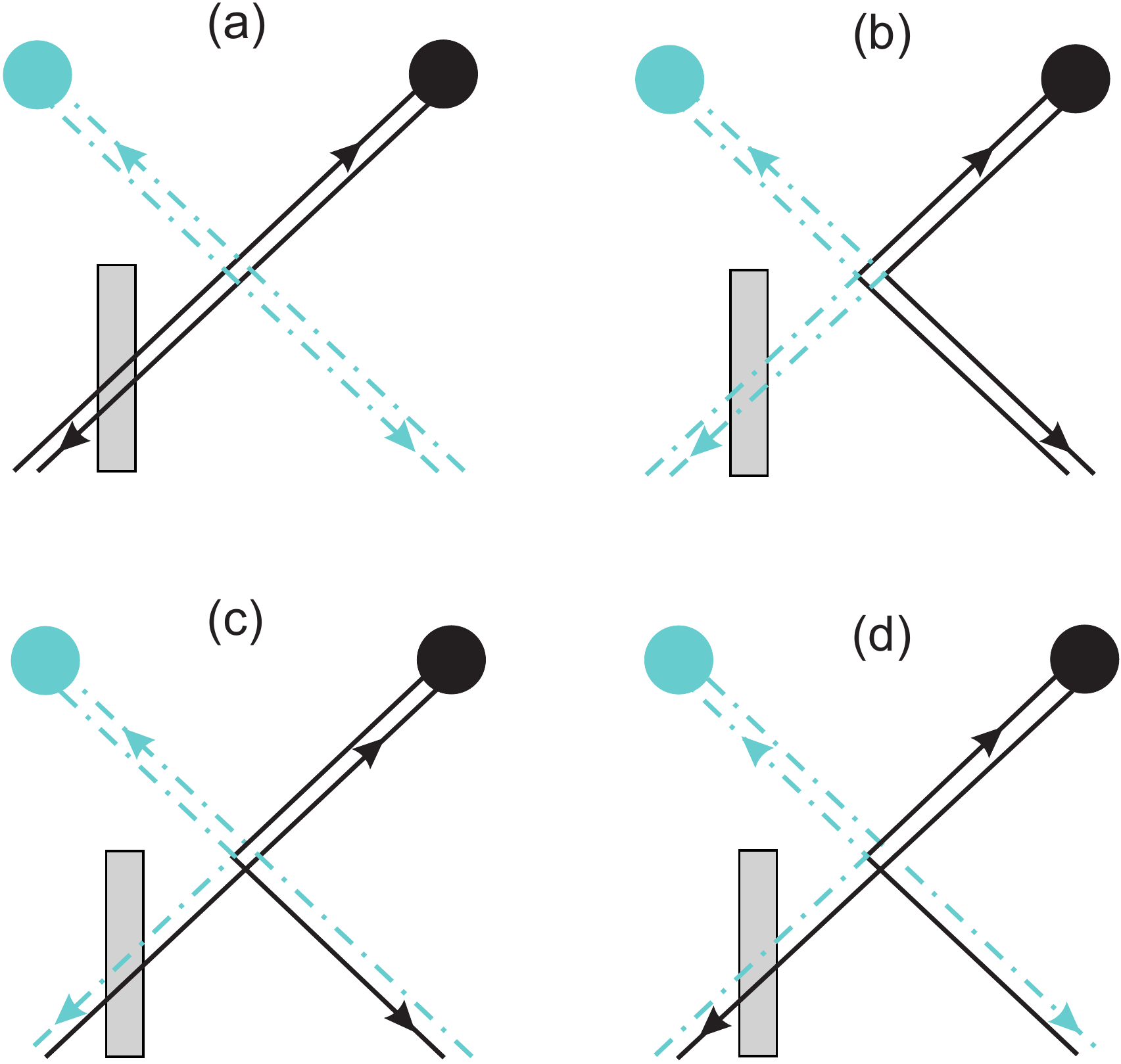}
 \caption{\label{fig:space-time} Color online.
 Space time diagrams for the probability $P_{+-}$: the vertical axis represent
 time whereas the horizontal one represent space.
 The blue and black dots (color online)
 represent destructive detection on the $+$ and $-$ output edge channels of the device.
 Trajectories connected to the dots have the same color (blue ones are also dashed-dotted
 for convenience).
 Upward arrows indicate forward branches in the Feynman-Vernon double path integral
 whereas downward arrows indicate backward branches.
 A straight trajectory means that the particle remains on the same track whereas a broken
 trajectory involves a jump from one track to the other at the QPC.
 The grey rectangle represents the action of our detector.
 Diagrams (a) and (b) correspond to the classical partition by the QPC whereas diagrams (c) and (d)
contain exchange effects associated with identical quantum particles.}
\end{figure}

%% Correction syntaxique
For classical distinguishable particles, the probability
$P_{+-}$ is the sum of two contributions: one in which particles
remain on the same track and one in which particles change track
while crossing the QPC. These correspond to space-time diagrams (a)
and (b) on Fig. \ref{fig:space-time}. They involve decoherence
coefficients along parallel trajectories that reach the same
detection point of space time. Thus, along these trajectories, the
distance $|(x^+-x^-)(\tau)|$ remains small (exactly zero for chiral
fermions). Therefore, these contributions are unaffected by the
presence of the detector.

For quantum identical particles, quantum statistics
enters the game. The initial quantum
state is a symmetrized or antisymmetrized two-particle state depending on
the bosonic or fermionic character of the particles. Therefore, the probability $P_{+-}$ also
contains quantum contributions that are represented by diagrams (c) and (d)
on Fig. \ref{fig:space-time}. In these contributions, the decoherence coefficient involves trajectories
that are not connected to the same terminal detection points in space time. Therefore
these contributions are much more sensitive to the effect of the gate since they feel the
difference in the positions of the detection points. In the limit of very strong
decoherence, the decoherence factor that weight contributions (c) and (d)
kills them, giving back the results that would be obtained for distinguishable particles.

Assuming that the two particles were initially injected with the same wave function
on the two input branches, the probabilities $P_{+-}$ and $P_{++}$ in the presence of the gate are
equal to:
\begin{eqnarray}
P_{+-}& = &\frac{1}{4} \int
|\psi(x)\,\psi(x')|^2 (1+e^{-\Re\big[\Phi_{FV}(x,x')\big]}
)\,dx\,dx' \nonumber \\
\\
P_{++}& = &\frac{1}{4} \int
|\psi(x)\,\psi(x')|^2 (1-e^{-\Re\big[\Phi_{FV}(x,x')\big]}
)\,dx\,dx'\,.
\nonumber \\
\end{eqnarray}
Thus, the two-particle interference can potentially probe the
decoherence coefficient associated with the gate. Recent experiments
have been able to observe electronic interference in a electronic
Mach-Zehnder set-up and coherence lengths of order 20~$\mu$m have
been reported~\cite{Roulleau07,Buttiker}. Visibility of fringes up
to 90~\% have been obtained~\cite{Heiblum07-2}.
%% One extra precision
The same device has also detected the signature of two particle interferences on current cross-correlations,
thus demonstrating quantum coherence along a length path of $2\times 8\ \mu$m. This suggests that
the manipulations envisioned in this paper could be performed in the near future.

%At this point, it is worth mentioning that in a realistic experimental setup, other effects may also contribute
%to a reduction of the antibunching effect. These include an incomplete
%matching of the wavefunctions produced by the detectors, any
%time variation of the state of the particles emitted, limitations on partial coherence of the initial wavepackets etc.
%Further work will be necessary to develop a realistic modeling of such a two edge excitation interference experiment
%but the simple model shown in the present paper provides the basic concepts needed to move forward.
%
%%%%%%%%%%%%%%%%%%%%%%%%%%%%%%%%%%%%%%%%%%%%%%%%%%%%%%%%%%%%%%%%%%%%%%%
\section{Conclusion and perspectives}
\label{sec:conclusion}

We have performed a detailed analysis of a solid-state device appropriate
for the detection of flying qubits at the edge of a two-dimensional electron gas
in the IQHE regime. A very special feature of this device is that the interaction with
the electron is restricted to a small interval of time. As a consequence,
it is no longer possible to discuss the quantum limit of measurement by comparison
of a measurement time and a decoherence time. However we have shown that, by
invoking information-theoretic concepts, the appropriate notion of
quantum limit can be defined. Using this approach, we have discussed in a very explicit way the deviation from
the quantum limit as well as the detector optimization, having in mind the
perspective of forthcoming experiments.

\medskip

Although the present work focuses on a very simple detector scheme,
our approach could in principle be used to deal with more realistic
situations.
%% Correction
A detection scheme based on a damped
oscillator treated as an LC circuit connected to a transmission line \cite{Johansson:2006-1}
could be treated using our method. But one could also think of using
a mesoscopic detector such as a nanotube transistor
\cite{Javey:2004} or a quantum point contact, the latter being known
for its ability to reach the quantum limit in a continuous
measurement \cite{Clerk03,Clerk06}. A more realistic description of
the preamplifier in the quantum regime is needed. Its backaction could
be accounted for within the present formalism assuming it generates
a Gaussian noise, possibly out of equilibrium. Note that
taking more precisely the added noise and the associated quantum constraints into account
would lead to a precise description of the output amplified signal.
Recently, measurement of the state of a qubit using a Josephson bifurcation amplifier
(JBA) has been performed \cite{Siddiqi:2004-1}. In this dispersive
measurement technique, the backaction noise arises from sources that
can be thermalized efficiently to the lowest temperature available.
Inventing and studying a non linear bifurcation amplifier able to
work in the presence of a high magnetic field would certainly be
interesting perspective for flying qubit detection.

%%%%%%%%%%%%%%%%%%%%%%%%%%%%%%%%%%%%%%%%%%%%%%%%%%%%%%%%%%%%%%%%%%%%%%%%%%%%%%%

\begin{acknowledgments}
We thank the Mesoscopic Physics group at LPA-ENS for constant
interaction during this work. P.D. acknowledges support from the
Condensed Matter Theory visitors program at Boston University.
\end{acknowledgments}

%%%%%%%%%%%%%%%%%%%%%%%%%%%%%%%%%%%%%%%%%%%%%%%%%%%%%%%%%%%%%%%%%%%%%%%%%%%%%%%
% Appendixes

\appendix
\section{Derivation of the Langevin equation}
\label{appendix:Langevin}

The idea is to compute the
effective action to the leading order in the quantum fluctuation
$\Delta(t)=x^+(t)-x^-(t)$ and to derive a semi-classical equation of motion for
the average position variable $r(t) = \frac{1}{2}({x^+(t)+x^-(t)})$.
In the high temperature regime, the effective action for the electron becomes local :
\begin{eqnarray}
\frac{iS}{\hbar} & = & i\frac{m}{ \hbar} \int dt \,\dot{r}\dot{\Delta}
-i\frac{e^{2}R}{ \hbar} \int dt  f'(r)^2 \,\dot{r}\,\Delta \nonumber \\
 & - & \frac{e^2R k_{B}T}{ \hbar^{2}} \int dt f'(r)^{2} \Delta^{2}
\end{eqnarray}
Considering the kernel \eqref{eq:electron:kernel} for $x^+_{f}=x^-_{f}=x_{f}$
and $x_{i}^+=x_{i}^-=x_{i}$,
integrating the first part and transforming the last
Gaussian term into a Gaussian integral over the auxiliary field $\xi$ leads to the
following path integral:
\begin{widetext}
\begin{eqnarray}
J_{t}\left(
\begin{array}{cc}
x_{f} & x_{i}\\
x_{f} & x_{i}
\end{array}
\right) & = &
\int \mathcal{D}[r,s,\xi]\, e^{-\int d\tau \frac{ \xi(\tau)^{2}}{4 e^{2} R k_{B}T}}
e^{i\hbar^{-1} \int d\tau \; \Delta(\tau) [-m \ddot{r}(\tau)-e^{2}R f'(r(\tau))^{2}
\dot{r}(\tau) +f'(r(\tau)) \xi(\tau)]} \\
& = & \int \mathcal{D}[r,\xi] \,e^{-\int d\tau \frac{
\xi(\tau)^{2}}{4 e^{2} R k_{B}T}}\prod_{\tau=0}^t\delta \big[ m
\ddot{r}(\tau) + e^{2}R f'(r(\tau))^{2} \dot{r}(\tau) -
f'(r(\tau))\xi(\tau)\big] \label{eq:master:langevin-path-integral}
\end{eqnarray}
\end{widetext}
where $r(0)=x_{i}$ and $r(t)=x_{f}$. The second line follows from the first by integration
over the quantum fluctuation $\Delta$. The r.h.s. of \eqref{eq:master:langevin-path-integral}
describes the Langevin equation \eqref{eq:quadratic:Langevin} and its associated noise
\eqref{eq:quadratic:Langevin:noise}.

\section{The decoherence factor for chiral fermions}
 \label{appendix:decoherence-factor}

 In full generality, the Feynman-Vernon influence functional for a pair of trajectories $[x^+(\tau),x^-(\tau)]$ is defined by
 tracing out over the RC circuit degrees of freedom. In operatorial form, it is given by:
 \begin{equation}
 \mathcal{F}[x^+,x^-]=\mathrm{Tr}\left(
 U_{I}[x^+,t]\ldotp \rho_{r,0}\ldotp U_{I}^\dagger[x^-,t]
 \right)
 \label{eq:FV:operator}
 \end{equation}
where $U_{I}[x^\pm,t]$ denotes the evolution operator between
$t_{i}=0$ and $t$ for the RC circuit in the
presence of an electron moving along $x^\pm(\tau)$:
\begin{equation}
U_{I}[x^\pm,t]= \mathrm{T}\, e^{-i\frac{e}{\hbar C} \int_{0}^{t}
f(x^\pm(\tau)) Q_I(\tau)\,d\tau} \,.
\end{equation}
In the case of chiral fermions, the electron dynamics can be solved exactly in the
interaction representation leading to Eq. \eqref{eq:evolution:chiral}.
Using \eqref{eq:FV:operator}, the decoherence coefficient given by Eq. \eqref{rhochir} is precisely equal to the Feynman-Vernon
influence functional for trajectories $x^\pm(\tau)=x_{i}^\pm+v_{F}\tau$. In
the case considered here, the Feynman-Vernon influence functional is a Gaussian functional of $\tau \mapsto
f(x^\pm(\tau))$ given by Eqs. \eqref{PhiFV} and \eqref{bigL}.

\section{Master equation for decoherence}
\label{appendix:master-equation}

In the high temperature limit, the evolution of the density matrix can be rewritten
as a master equation:
\begin{eqnarray}
\frac{\partial \rho_{el}(t)}{\partial t} & = & \frac{i
\hbar}{2m}\big[\frac{\partial ^{2}}{\partial x^{2}} - \frac{\partial
^{2}}{\partial x'^{2}}\big] \rho_{el}
\label{eq:decoherence:master-eq:free} \\
 & - & \frac{e^{2}R}{2m}\big[f(x)-f(x')\big]\big[f'(x)\frac{\partial}{\partial
x } -f'(x')\frac{\partial}{\partial x' }\big]\rho_{el} \nonumber \\
\label{eq:decoherence:master-eq:friction} \\
 &  - & \frac{e^{2}Rk_{B}T}{ \hbar^{2}}
 \big[f(x)-f(x')\big]^{2}\rho_{el}
\label{eq:decoherence:master-eq:decoh}
\end{eqnarray}
Assuming the energy released by the edge is weak, we shall
neglect the friction term \eqref{eq:decoherence:master-eq:friction} and focus
on the decoherence term \eqref{eq:decoherence:master-eq:decoh}. The resulting simplified equation
rewrites:
\begin{eqnarray}
\frac{\partial \rho_{el}(t)}{\partial t} & = & \frac{i
\hbar}{2m}\left(\frac{\partial ^{2}}{\partial x^{2}} - \frac{\partial
^{2}}{\partial x'^{2}}\right) \rho_{el}(t)\nonumber \\
& - & \frac{e^{2}R k_{B}T}{ \hbar^{2}}
\big[f(x)-f(x')\big]^{2}\rho_{el}(t)
\label{eq:decoherence:master:simplified}
\end{eqnarray}
We will look for a solution of \eqref{eq:decoherence:master:simplified} in the following
form:
\begin{equation}
\rho_{el}(t) = \rho_{\mathrm{free}}(t)\times (\delta \rho)(t)
\end{equation}
Where $\rho_{\mathrm{free}}(t)$ describes the free evolution of the particles~:
\begin{eqnarray}
\langle x|\rho_{\mathrm{free}}(t)|x'\rangle & \sim & e^{i
k_F(x-x')}e^{-\frac{(x-v_Ft)^{2}}{a^{2}+2i\hbar
t/m}}e^{-\frac{(x'-v_Ft)^{2}}{a^{2}+2i\hbar t/m}} \nonumber \\
\end{eqnarray}
The evolution equation of $\delta \rho$ contains terms of the same form as the r.h.s. of
\eqref{eq:decoherence:master:simplified} plus
a contribution arising from single derivatives of
$\rho_{\mathrm{free}}(x,x',t)$ with respect to $x$ and $x'$.
Assuming that the smallest length scale in the problem is $\lambda_{F}=2\pi/k_{F}$,
the derivatives of $\rho_{\mathrm{free}}(x,x',t)$ can be approximated by:
\begin{eqnarray}
\frac{\partial \rho_{\mathrm{free}}}{\partial x} & \simeq &
ik_{F}\,\rho_{\mathrm{free}}(x,x',t)\\
\frac{\partial \rho_{\mathrm{free}}}{\partial x'} & \simeq &
-ik_{F}\,\rho_{\mathrm{free}}(x,x',t)
\end{eqnarray}
The resulting contribution then dominates the kinetic energy term in the evolution
equation for $\delta\rho$. With these approximations, the evolution of $\delta \rho$ is then given by:
\begin{eqnarray}
\frac{\partial (\delta \rho)}{\partial t} & = & -v_{F}\left(\frac{\partial
}{\partial x} + \frac{\partial }{\partial x'}\right)\,(\delta \rho)\nonumber \\
& - & \frac{e^{2}R k_{B}T}{ \hbar^{2}} (f(x)-f(x'))^{2}(\delta \rho)
\label{eqdrho}
\end{eqnarray}
whose solution is given by
\begin{equation}
(\delta \rho) (x_{f}^+,x_{f}^-,t)   =  e^{-\frac{e^{2}R k_{B}T}{ \hbar^{2}}\int_{0}^{t}
(\Delta_{f}(\tau))^2\,d\tau }
\end{equation}
where $x_{i}^\pm=x_{f}^\pm-v_{F}t$ and $\Delta_{f}(\tau)=f(x_{f}^+-v_{F}\tau)-
f(x_{f}^--v_{F}\tau)$.
The effect of the interaction with the gate on the off-diagonal
elements of the density matrix is the same as the one found in the
high temperature limit for chiral fermions (see Eqs. \eqref{eq:decoherence:high-T}
and \eqref{eq:decoherence:high-T:aux}).
Indeed, Eq. \eqref{eqdrho} is identical to the one expected for a
linear dispersion relation $\epsilon_{k}=\hbar v_Fk$. Keeping the
leading terms of order $k_F$, we have neglected the small
fluctuations of the velocity $\delta k<<k_F$ around its mean value
and we have taken $k\approx k_F$. This approximation is valid if
the width of the wavepacket $\delta x$ is much larger than the fermi
wavelength $\lambda_F$ which means that we have a well defined
excitation above the Fermi energy: $\delta
\epsilon<<\epsilon_{f}$.

%%%%%%%%%%%%%%%%%%%%%%%%%%%%%%%%%%%%%%%%%%%%%%%%%%%%%%%%%%%%%%%%%%%%%%%%%%%%%%%%

\end{document}